\documentclass{LMCS}
\usepackage{enumerate,hyperref,breakurl}

\usepackage[latin9]{inputenc}
\usepackage[longnamesfirst]{natbib}

\usepackage{upgreek}
\usepackage[all]{xy}
\allowdisplaybreaks[4]
\newcommand{\trans}[2]{
\newenvironment{#1}{\begin{#2}}{\end{#2}}}

\trans{proposition}{prop}
\trans{lemma}{lem}
\trans{corollary}{cor}
\trans{theorem}{thm}
\trans{definition}{defi}
\trans{example}{exmp}
\trans{conjecture}{conj}
\trans{note}{rem}

\usepackage{wasysym}

\newcommand{\rien}[1]{}
\renewcommand{\phi}{\varphi}

\newcommand{\rulesCaption}{}
\newcommand{\rulesLabel}{}

\newcommand{\vect}[2][n]{#2_{1},\ldots,#2_{#1}}

\newcommand{\rewrite}{\mathop{\longrightarrow}\limits}
\newcommand{\sRew}{\mathop{\longleftrightarrow}\limits}

\def\terms#1#2{\mathcal T(#1,#2)}

\newcommand{\subterm}[2]{{#1}_{|#2}}
\newcommand{\replacement}[3]{{#1}[#2]_{#3}}

\catcode`\«=\active
\catcode`\»=\active
\def«{\og\ignorespaces}
\def»{{\fg}}

\newcommand{\subst}[3]{\{#2/#3\}#1}

\newcommand{\longv}[1]{}

\providecommand{\Def}[1]{{\em #1}}
\providecommand{\df}{~~~\stackrel{\text{def}}=~~~}

\let\imp = \Rightarrow 

\newcommand{\theory}{\mathcal}
\usepackage{bussproofs}
\def\fCenter{\vdash}

\def\Tndd$#1${\AxiomC{}\LeftLabel{\emph{classical}}
\UnaryInfC{$#1$}}
\def\Tnda[#1]$#2${\AxiomC{}\LeftLabel{\emph{classical}}\RightLabel{\scriptsize#1}
\UnaryInfC{$#2$}}
\def\ImpId$#1${\LeftLabel{$\imp$-i}
\UnaryInfC{$#1$}}
\def\ImpIa[#1]$#2${\RightLabel{\scriptsize#1}\ImpI$#2$}
\def\ImpEd$#1${\LeftLabel{$\imp$-e}
\BinaryInfC{$#1$}}
\def\ImpEa[#1]$#2${\RightLabel{\scriptsize#1}\ImpE$#2$}
\def\AndId$#1${\LeftLabel{$\wedge$-i}
\BinaryInfC{$#1$}}
\def\AndIa[#1]$#2${\RightLabel{\scriptsize#1}\AndI$#2$}
\def\AndEd$#1${\LeftLabel{$\wedge$-e}
\UnaryInfC{$#1$}}
\def\AndEa[#1]$#2${\RightLabel{\scriptsize#1}\AndE$#2$}
\def\OrId$#1${\LeftLabel{$\vee$-i}
\UnaryInfC{$#1$}}
\def\OrIa[#1]$#2${\RightLabel{\scriptsize#1}\OrI$#2$}
\def\OrEd$#1${\LeftLabel{$\vee$-e}
\TrinaryInfC{$#1$}}
\def\OrEa[#1]$#2${\RightLabel{\scriptsize#1}\OrE$#2$}
\def\AllId$#1${\LeftLabel{$\forall$-i}
\UnaryInfC{$#1$}}
\def\AllIa[#1]$#2${\RightLabel{\scriptsize#1}\AllI$#2$}
\def\AllEd$#1${\LeftLabel{$\forall$-e}
\UnaryInfC{$#1$}}
\def\AllEa[#1]$#2${\RightLabel{\scriptsize#1}\AllE$#2$}
\def\ExId$#1${\LeftLabel{$\exists$-i}
\UnaryInfC{$#1$}}
\def\ExIa[#1]$#2${\RightLabel{\scriptsize#1}\ExI$#2$}
\def\ExEd$#1${\LeftLabel{$\exists$-e}
\BinaryInfC{$#1$}}
\def\ExEa[#1]$#2${\RightLabel{\scriptsize#1}\ExE$#2$}

\def\BotEd$#1${\LeftLabel{$\bot$-e}
\UnaryInfC{$#1$}}
\def\BotEa[#1]$#2${\RightLabel{\scriptsize#1}\BotE$#2$}
\def\TopId$#1${\AxiomC{}\LeftLabel{$\top$-i}
\UnaryInfC{$#1$}}
\def\TopIa[#1]$#2${\RightLabel{\scriptsize#1}\TopI$#2$}

\def\atocd$#1$$#2${\AxiomC{[$#1$]}\noLine\UnaryInfC{$#2$}}
\def\atoca[#1]$#2$$#3${\AxiomC{[$#2$]}\noLine\LeftLabel{$#1$\{}\UnaryInfC{$#3$}}

\def\Axiomm[#1]$#2${\AxiomC{$#2${ \scriptsize (#1)}}}

\makeatletter
\def\beamerify#1{%
\expandafter\def\csname #1ba\endcsname<##1>[##2]$##3${\csname #1a\endcsname[\onslide<##1>{##2}]$\onslide<##1>{##3}$}
\expandafter\def\csname #1bd\endcsname<##1>$##2${\csname #1d\endcsname$\onslide<##1>##2$}
\expandafter\def\csname #1b\endcsname<##1>{\@ifnextchar[{\csname #1ba\endcsname<##1>}{\csname#1bd\endcsname<##1>}}
\expandafter\def\csname #1n\endcsname{\@ifnextchar[{\csname #1a\endcsname}{\csname #1d\endcsname}}
\expandafter\def\csname #1\endcsname{\@ifnextchar<{\csname #1b\endcsname}{\csname #1n\endcsname}}
}
\makeatother

\makeatletter
\def\Tnd{\@ifnextchar[{\Tnda}{\Tndd}}
\def\ImpI{\@ifnextchar[{\ImpIa}{\ImpId}}
\def\AndI{\@ifnextchar[{\AndIa}{\AndId}}
\def\OrI{\@ifnextchar[{\OrIa}{\OrId}}
\def\AllI{\@ifnextchar[{\AllIa}{\AllId}}
\def\ExI{\@ifnextchar[{\ExIa}{\ExId}}
\def\ImpE{\@ifnextchar[{\ImpEa}{\ImpEd}}
\def\AndE{\@ifnextchar[{\AndEa}{\AndEd}}
\def\OrE{\@ifnextchar[{\OrEa}{\OrEd}}
\def\AllE{\@ifnextchar[{\AllEa}{\AllEd}}
\def\ExE{\@ifnextchar[{\ExEa}{\ExEd}}
\def\BotE{\@ifnextchar[{\BotEa}{\BotEd}}
\def\TopI{\@ifnextchar[{\TopIa}{\TopId}}
\def\atoc{\@ifnextchar[{\atoca}{\atocd}}
\makeatother

\def\ifundefined#1{\expandafter\ifx\csname#1\endcsname\relax}

\newcommand\lvdash[2][]{\mathrel{\rule{.08ex}{1.5ex}\kern-.12em\lower-.08em\hbox{$\frac{\scriptstyle\,#1\hfill}{\scriptstyle~#2\,}$}}}
\newcommand\Hvdash{\lvdash[\textup{\textsf{S}}]}
\newcommand\Nvdash{\lvdash[\textup{\textsf{N}}\!]}

\providecommand{\imp}{\Rightarrow}
\providecommand{\ra}{\rightarrow}
\newcommand{\rewsys}{\mathcal}
\newcommand{\R}{\rewsys R}
\newcommand{\WS}{\rewsys{WS}}
\newcommand{\HO}{\rewsys{HO}}
\newcommand{\HHA}{\rewsys{HHA}^{\text{mod}}}
\newcommand{\bigO}{O}
\newcommand{\ist}{$^{\text{st}}$}
\newcommand{\ith}{$^{\text{th}}$}
\newcommand{\T}[2][]{T$_{#1}\left(\text{#2}\right)$}
\newcommand{\Tm}[2][]{\ensuremath{\textrm{T}_{#1}\left(#2\right)}}
\newcommand{\MP}{\LeftLabel{(\ref{eq:mp})}\BinaryInfC}
 
\def\doi{7 (1:3) 2011}
\lmcsheading%
{\doi}
{1--31}
{}
{}
{Apr.~15, 2010}
{Mar.~16, 2011}
{}

\begin{document}

\title[Efficiently Simulating Higher-Order Arithmetic by a \texorpdfstring{1\ist}{1st}-Order
  Theory Modulo]{Efficiently Simulating Higher-Order Arithmetic\\ by a First-Order
  Theory Modulo\rsuper*}
\author[G.~Burel]{Guillaume Burel}
\address{Max Planck Institute for Informatics\\
Saarland University\\
Saarbrücken, Germany}
\email{guillaume.burel@ens-lyon.org}
\thanks{Present address: École Nationale Supérieure d'Informatique pour l'Industrie et l'Entreprise,
1 square de la résistance,
91025 Evry Cedex, France}

\newcommand{\sep}{,\ } 


\titlecomment{{\lsuper*}Parts of this paper have previously appeared in
  \citep{burel07csl}. In particular, this paper contains the proofs
  missing there.}

\begin{abstract}
  In deduction modulo, a theory is not represented by a set of axioms
  but by a congruence on propositions modulo which the inference rules
  of standard deductive systems---such as for instance natural
  deduction---are applied. Therefore, the reasoning that is intrinsic
  of the theory does not appear in the length of proofs. In general,
  the congruence is defined through a rewrite system over terms and
  propositions.  We define a rigorous framework to study proof lengths
  in deduction modulo, where the congruence must be computed in
  polynomial time. We show that even very simple rewrite systems lead
  to arbitrary proof-length speed-ups in deduction modulo, compared to
  using axioms. As higher-order logic can be encoded as a first-order
  theory in deduction modulo, we also study how to reinterpret, thanks
  to deduction modulo, the speed-ups between higher-order and
  first-order arithmetics that were stated by Gödel. We define a
  first-order rewrite system with a congruence decidable in polynomial
  time such that proofs of higher-order arithmetic can be linearly
  translated into first-order arithmetic modulo that system. We also
  present the whole higher-order arithmetic as a first-order system
  without resorting to any axiom, where proofs have the same length as
  in the axiomatic presentation.
\end{abstract}

\maketitle
\section{Introduction}
Studying the length of the proofs produced by a logical system can of
course have practical motivations. Indeed, shorter
proofs seem to be easier to find out---either by hand or
automatically---, to share and to maintain. Automated provers may be
able to find proofs that are longer than proofs done by humans, they
have nevertheless bounded capacities. Even if computing power is
always increasing, so that one is no longer afraid to use SAT-solvers
within verification tools (mainly because worst cases do not often
occur in practice), it is not conceivable to build an automated
theorem prover that produces only proofs of non-elementary length.

This study is also theoretically interesting. As remarked by Parikh in
the introductory paragraph of \citet{Godel:86}, ``the celebrated P=NP?
question can itself be thought of as a speed-up question.''
(See also \citealp{cook79relative}.) This explains the research for
speed-ups between proof systems---for instance, it is shown that Frege
systems have an exponential speed-up over resolution for propositional
logic~\citep{buss87pigeonhole}---and for new formalisms whose
deductive systems provide smaller proofs, such as for instance the
calculus of structures of~\citet{BruTh} w.r.t. the sequent calculus
of~\citet{gentzen:untersuchungen}
\citep[see][]{bruscoli09complexity}. The goal is to find a so-called
super proof system \citep{cook74lengths}, which can build polynomially
sized proofs of each propositional tautology, or to refute the
existence of such a system, in which case NP$\neq$coNP, which would
imply P$\neq$NP.  In this paper, the length of a proof corresponds to
its number of steps (sometimes called lines), whatever the actual size
of the propositions\ appearing in them is. 

Proofs are rarely searched for without
context:  mathematical proofs rely on set theory, or Euclidean
geometry, or arithmetic, etc.; proofs of program correctness are done
using e.g.\ pointer arithmetic and/or theories defining data structures
(chained lists, trees, \dots); concerning security, theories are used for
instance to model properties of encryption algorithms. In this paper,
we are therefore interested in the length of proofs in a theory. This
length may depend on several factors. First, the strength of the
theory plays a key role, as shown by the following result: it has been
proved by \citet{parikh73length} that second-order arithmetic
provides shorter proofs than first-order arithmetic. (This result was
stated earlier by \citet{goedel36laenge}, unfortunately without
proof.)\enspace This was generalized to all orders by~\citet{krajicek}, and
was proved for the true language of arithmetic by~\citet{buss94goedel}.
(The former results used an axiomatization of arithmetic using ternary
predicates to represent addition and multiplication.)
The theorem
proved by Buss is stated as follows:
  \begin{theorem}[{\citet[Theorem 3]{buss94goedel}}]\label{theo:buss}\hypertarget{hl:theo:buss}{}
   Let $i\geq 0$. There is an infinite family $\mathcal F$ of
   propositions of the language of first-order arithmetic such that 
   \begin{enumerate}[\em(1)]
   \item for all $P\in\mathcal F$, $Z_i\vdash P$
   \item there is a fixed $k\in\mathbb N$ such that for all
     $ P\in\mathcal F$, $Z_{i+1}\lvdash{k\text{ steps}} P$
   \item there is no fixed $k\in\mathbb N$ such that for all
     $ P\in\mathcal F$, ${Z_i\lvdash{k\text{ steps}} P}$. \qed
   \end{enumerate}
  \end{theorem}
\noindent where $Z_i$ corresponds to the $(i+1)$\ith-order arithmetic (so $Z_0$
is in fact first-order arithmetic), and ${Z_i\lvdash{k\text{
      steps}}P}$ means that $P$ can be proved in at most $k$ steps
within a schematic system ---i.e.\ a Hilbert-type (or Frege) system
with a finite number of axiom schemata and inference rules--- for
$(i+1)$\ith-order arithmetic. (In fact, Buss proved this theorem also for
weakly schematic systems, i.e.\ schematic systems in which every
tautology can be used as an axiom, as well as generalizations of
axioms, but we will not use this fact here.)

The length of the proofs depends also on the presentation of the
theory. For instance, if we present the theory $\theory T$ by taking all the
propositions that hold in that theory (${\{P: \theory T\models P\}}$)
as axioms,
then for all true propositions $P$ there is a one-step ``proof'',
namely using the axiom $P$. Of course, we can argue whether those are
really proofs. Indeed, in that case, proof checking consists of
checking that $P$ holds in $\theory T$, and is therefore undecidable
in general. On the other hand, using a finite first-order
axiomatization of the theory does not seem optimal, in particular when
computations are involved. For instance, a proof of $2+2=4$ should be
straightforward and should not contain more than one step that
consists of checking the computation that ``$2+2$ makes $4$''. Then, it seems
important to distinguish what part of a proof corresponds to
computation and what part is real deduction, in order to better combine
them. Such an idea is referred to as Poincaré's principle.  Deduction
modulo \citep{DHK-TPM-JAR} is a formalism deriving from this
principle. The computational part of a proof is put in a congruence
between propositions\ modulo which the application of the deduction
rules takes place. This leads for instance to the sequent calculus
modulo and to the natural deduction modulo. The congruence is often
defined as a set of rewrite rules that can rewrite terms but also
\emph{atomic propositions}. Indeed, one wants for instance to consider
the definition of the addition or multiplication using rewrite rules
over terms as part of the computation, but also the following rewrite
rule:
$$x\times y = 0~\ra~ x = 0\vee y = 0\enspace.$$ This rule rewrites an atomic
proposition to a proposition. Then, the following simple
natural-deduction-modulo proof of $t\times t = 0$ can be deduced from
a proof $\pi$ of $t=0$:
$$
  \AxiomC{$\pi$}
  \noLine\UnaryInfC{$t=0$}
  \OrI[$t\times t=0~\rewrite~ t = 0\vee t = 0$]$t\times t=0$
  \DisplayProof\enspace.
$$
{\sloppy
Rewriting of propositions is essential to being able to encode expressive
theories in \break deduction modulo, as has been done for first-order
arithmetic~\citep{arith}, Zermelo's set theory~\citep{DowekMiquel},
simple type theory (a.k.a.\ higher-order logic)~\citep{hollambdasigma}
or pure type systems~\citep{CousineauDowek,burel08pts}.
}

As computations are not part of the deduction in the proof, they
should not be counted in the length of the proof. Indeed, a proof in
deduction modulo consists only of the deductive steps, and the
computational steps are replayed during proof checking. However, this
is too general if we are concerned with the notion of proof
length. Because rewriting is Turing-complete, a whole proof system can
be encoded in the computational part. This leads to the same problem
as using all propositions of the theory as axioms: proof checking is
no longer decidable. We therefore need a more rigorous framework to
study proof length in deduction modulo.  We argue that we should only
call a proof an object that can be checked feasibly, that is, in
polynomial time. This is of course an arbitrary criterion (we could
for instance have chosen another complexity class), but it seems
natural.  Furthermore, this is requested if one wants to link proof
theory with complexity theory. Indeed, \citet{cook79relative} defined a
framework in which a proof system for a theory $\theory T$ is an onto
function computable in polynomial time from the words over some
alphabet (representing the proofs) to the set of propositions that
hold in $\theory T$. Starting from a more conventional proof system,
the idea is to map a correct proof with its conclusion, and an
incorrect proof to any proposition of $\theory T$. As the function
must be computable in polynomial time, proof checking in the real
system has to be feasible. In deduction modulo, this requirement
implies that the congruence must be checkable in polynomial time. In
this paper, we will consider rewrite systems that are confluent and
that have a polynomial derivational complexity, i.e.\ the number of
rewrite steps of a term of size $n$ must be bounded by a polynomial of
$n$.

Deduction modulo is logically equivalent to the axiomatic theory
corresponding to the congruence~\citep[Proposition 1.8]{DHK-TPM-JAR},
but proofs are often considered as simpler, because the computation is
hidden, letting the deduction clearly appear. Proofs are also claimed
to be shorter for the same reason. Nevertheless, this fact was never
quantified. Besides, it is possible, in deduction modulo, to build proofs of
Higher-Order Logic using a first-order
system~\citep{hollambdasigma}. Using this, a step of higher-order
resolution is completely simulated by a step of ENAR, the resolution
and narrowing method based on deduction modulo. It looks like this is
also the case for the associated sequent calculi, although this was not
clearly stated.  Therefore, it seems reasonable to think that
deduction modulo is able to give the same proof-length speed-ups as
the ones occurring between $(i+1)$\ith- and $i$\ith-order
arithmetic. This paper therefore investigates how to relate
proof-length speed-ups in arithmetic with the computational content of
the proofs.

Our first result is to show that even a very simple rewrite system can
lead to arbitrary proof-length
speed-ups~(Theorem~\ref{theo:simple}). By arbitrary proof-length
speed-up, we mean, as in Theorem~\ref{theo:buss}, that we can find a
family of propositions that can be proved by a bounded number of steps
in one system, whereas in the other, the minimal proof length depends
on the proposition that is proved. Thus, proofs in the second system
are arbitrarily longer than in the first.  Then, we show how to encode
everything concerning higher orders up to $i>0$ into a confluent
rewrite system $\HO_i$ with polynomial derivational complexity. Modulo
this rewrite system, we show that it is possible to stay in
first-order arithmetic while preserving the proof lengths of
higher-order arithmetic (Theorem \ref{theo:s3}). This shows that the
origin of the speed-up theorem of Buss can be, at least to some
extend, expressed as simple computations. Note that $\HO_i$ is not the
restriction of the encoding of HOL by \citet{hollambdasigma} up to the
order $i$, because we were not able to prove that its derivational
complexity is bounded.

In this paper, we are also concerned with extending the work of
\citet{arith}, in which the whole first-order arithmetic is expressed
as a rewrite system. In that case, we speak of a purely computational
presentation of the theory. Thus, we show how to express
\emph{higher-order} arithmetic as a purely computational theory. This
permits to recover desirable properties such as disjunction and
witness properties for higher-order Heyting arithmetic (i.e.\
intuitionistic arithmetic). This is not just the combination of the
encoding of higher orders and the formulation of first-order
arithmetic by \citet{arith}, because the latter does not preserve the
length of proofs.  We define higher-order arithmetic as a purely
computational theory $\HHA_i$ which has the same speed-up over
first-order arithmetic as the axiomatic presentation. Unfortunately,
the rewrite system of this presentation is not terminating. The rule
that breaks the termination is the one encoding the induction
principle, which is not surprising, because this is where the strength
of arithmetic lies. We therefore advocate the use of a new inference
rule corresponding to it.

This works revisits and extends a previous work \citep{burel07csl}
where we looked at the relations between computations and proof-length
speed-ups. We work in a much more rigorous framework here. For
instance, in \citeyear{burel07csl}, we only stated that the rewrite
systems we were using were ``simple'', whereas we request here that
they are confluent and with a polynomially bounded derivational
complexity. Also, in \citeyear{burel07csl}, in the translation of
$Z_i$ to $Z_{i-1}$ modulo, there remained axioms in which function
symbols of order $i$ were involved, which is no longer the case here.

The next section will present the minimal knowledge needed on
deduction modulo to make the paper self-contained, it defines the
notion of polynomially bounded derivational complexity, and shows that
arbitrary proof-length speed-ups naturally occur thanks to deduction
modulo, even for very simple rewrite systems with polynomially bounded
derivational complexity. In Section~\ref{sec:arith} we present proof
systems for higher-order arithmetic, and we prove that using schematic
systems or natural deduction is not relevant w.r.t.\ arbitrary
proof-length speed-ups.  Then, Section~\ref{sec:main} 
presents how to efficiently encode higher orders, and then higher-order
arithmetic. Finally, in Section~\ref{sec:demo} we apply these
results to investigate the origin of the speed-ups in arithmetic.

\section{Proof Speed-ups in Deduction Modulo}
\subsection{Rewriting propositions}
In this section, we recall the definition of deduction modulo, as
introduced by \citet*{DHK-TPM-JAR} and \citet{normalization}.  In
deduction modulo, propositions\ are considered modulo some congruence
defined by some rules that rewrite not only terms but also
propositions. We use standard definitions, as given by \citet{allThat},
and extend them to proposition rewriting \citep{DHK-TPM-JAR}.

First, let us recall how to build many-sorted first-order propositions\ 
\citep[see][Chapter 10]{GallierLivreRevised}, mainly to
introduce the notations we will use.  A (first-order) many-sorted
signature consists of a set of function symbols and a set of
predicates, all of them with their arity (and co-arity for function
symbols).  We denote by $\terms\Sigma V$ the set of \Def{terms} built
from a signature $\Sigma$ and a set of variables $V$. An \Def{atomic
  proposition} is given by a predicate symbol $A$ of arity $[\vect i]$
and by $n$ terms $\vect t\in\terms\Sigma V$ with matching sorts. It is
denoted $A(\vect t)$.  \Def{Propositions} can be built using the
following grammar:
$$\mathcal P~::=~\bot~|~\top~|~A~|~\mathcal P\wedge \mathcal P~|~\mathcal
P\vee \mathcal P~|~\mathcal P\imp \mathcal P~|~\forall x.~\mathcal
P~|~\exists x.~\mathcal P$$ where $A$ ranges over atomic propositions
and $x$ over variables.  $P\Leftrightarrow Q$ is used as a syntactic
sugar for ${(P\imp Q)\wedge(Q\imp P)}$, as well as $\neg P$ for
$P\imp\bot$.  Positions in a term or a proposition, free variables and
substitutions are defined as usual \citep[see][]{allThat}. The
replacement of a variable $x$ by a term $t$ in a proposition $P$ is
denoted by $\subst Ptx$, the subterm or subproposition of $t$ at the
position $\mathfrak p$ by $\subterm t{\mathfrak p}$, and its
replacement in $t$ by a term or proposition $s$ by $\replacement
ts{\mathfrak p}$. Propositions are considered modulo
$\alpha$-conversion of the variables bound by $\forall$ and $\exists$.
Applying a substitution does not
capture variables: $\{s/x\}(P(x)\wedge \forall x.~P(x)) =
P(s)\wedge\forall x.~P(x)$. Replacing a subterm by another can capture
variables $(\forall x. P(x,t))[s(x)]_{1.2}=\forall
x.~P(x,s(x))\enspace$.

A \Def{term rewrite rule} is the pair of terms $l,r$ such that all
free variables of $r$ appear in $l$. It is denoted $l\rightarrow r$. A
\Def{term rewrite system} is a set of term rewrite rules.  A term $s$
can be rewritten to a term $t$ by a term rewrite rule $l\rightarrow r$
if there exists some substitution $\sigma$ and some position
$\mathfrak p$ in $s$ such that $\sigma l=\subterm s{\mathfrak p}$ and
$t = \replacement s{\sigma r}{\mathfrak p}$.  We extend this notion to
propositions: a proposition  $Q$ can be
rewritten to a proposition $R$ by a term rewrite rule $l\rightarrow r$
if there exists some substitution $\sigma$ and some position
$\mathfrak p$ in $Q$ such that $\sigma l=\subterm Q{\mathfrak p}$ and
$R = \replacement Q{\sigma r}{\mathfrak p}$.

A \Def{proposition rewrite rule} is the pair of an atomic proposition
$A$ and a proposition $P$, such that all free variables of $P$ appear in
$A$. It is denoted $A\rightarrow P$. A \Def{proposition rewrite
  system} is a set of proposition rewrite rules.  A proposition $Q$ can be
rewritten to a proposition $R$ by a proposition rewrite rule $A\rightarrow
P$ if there exists some substitution $\sigma$ and some position
$\mathfrak p$ in $Q$ such that $\sigma A=\subterm Q{\mathfrak p}$ and
$R =\replacement Q{\sigma P}{\mathfrak p}$. Semantically, this
proposition rewrite relation must be seen as a logical equivalence
between propositions.

A \Def{rewrite system} is the union of a term rewrite system and a
proposition rewrite system. The fact that $P$ can be rewritten to $Q$
either by a term or by a proposition rewrite rule of a rewrite system
$\R$ will be denoted by $A\rewrite_\R P$. The transitive
(resp. reflexive transitive) closure of this relation will be denoted
by $\rewrite^*_\R$ (resp. $\sRew^*_\R$).

\begin{definition}
  The \Def{derivational length} of a term or proposition $t$ w.r.t.\ a
  rewrite system $\rewsys R$ is the maximal length of a derivation
  starting from $t$ using $\rewsys R$. The \Def{derivational
    complexity} of a rewrite system $\rewsys R$ is the function that
  maps a natural number $n$ to the maximal derivational length w.r.t.\
  $\rewsys R$ of the terms and propositions of size at most $n$. 
\end{definition}
In
  this paper, we are interested in rewrite systems that are confluent
  and whose derivational complexity can be bounded by a
  polynomial. This implies that $\sRew^*_{\rewsys R}$ is decidable in
  polynomial time.
\subsection{Natural deduction modulo}
Using an equivalence $\sRew^*_\R$ defined by a term and proposition
rewrite system $\R$, we can define natural deduction modulo $\R$ as
\citeauthor{normalization} do \citeyearpar{normalization}. Its
inference rules are represented in Figure~\ref{fig:natmod}. They are
the same as the ones introduced by \citet{gentzen:untersuchungen},
except that we work modulo the rewrite relation $\sRew^*_\R$. Leaves
of a proof that are not discarded by the inference rules of the
proof (on the contrary to $A$ in $\imp$-i for instance) are the
assumptions of the proof. A cut in a proof is an introduction rule
immediately followed by an elimination rule. In particular, one says
that the proof cuts trough $A$ if there is a derivation
\begin{prooftree}
  \atoc$A$$B$ \ImpI$A\imp B$ \AxiomC{$A$} \ImpE$B$
\end{prooftree}
in it.

\begin{figure*}
\def\fCenter{\vdash_\R}
\begin{center}
  \begin{tabular}{@{}cc@{}}
    \atoc$A$$B$
    \ImpI[\ if $C\sRew_\R^*A\imp B$]$C$
    \DisplayProof
    &
    \AxiomC{$A$}
    \AxiomC{$C$}
    \ImpE[\ if $C\sRew_\R^*A\imp B$]$B$
    \DisplayProof
    \\[2em]
    \AxiomC{$A$}
    \AxiomC{$B$}
    \AndI[\ if $C\sRew_\R^*A\wedge B$]$C$
    \DisplayProof
    &
    \AxiomC{$C$}
    \AndE[\ if $C\sRew_\R^*A\wedge B$ or $C\sRew_\R^*B\wedge A$]$A$
    \DisplayProof
    \\[2em]
    \AxiomC{$ A$}
    \OrI[\ if $C\sRew_\R^*A\vee B$ or $C\sRew_\R^*B\vee A$]$ C$
    \DisplayProof
    &
    \AxiomC{$ C$}
    \atoc$A$$D$
    \atoc$B$$D$
    \OrE[\ if $C\sRew_\R^*A\vee B$]$ D$
    \DisplayProof
    \\[2em]
    \AxiomC{$\subst Ayx$}
    \AllI[\ \parbox{4.1cm}{if $B\sRew_\R^*\forall x.~A$ and $y$ is not
        free in $A$ nor in the
      assumptions of the proof above}]$ B$
    \DisplayProof
    &
    \AxiomC{$ A$}
    \AllE[\ if $A\sRew_\R^*\forall x.~C$ and $B\sRew_\R^*\subst Ctx$]$ B$
    \DisplayProof
    \\[2em]
    \AxiomC{$ B$}
    \ExI[\ if $A\sRew_\R^*\exists x.~C$ and $B\sRew_\R^*\subst Ctx$]$ A$
    \DisplayProof
    &
    \AxiomC{$ B$}
    \atoc$\subst Ayx$$\quad C\quad$
    \ExE[\ \parbox{4.1 cm}
     {if $B\sRew_\R^*\exists x.~A$ and $y$ is not free in $C$ nor
      in the assumption of the proof above except $\subst Ayx$}]$C$
    \DisplayProof
    \\[3em] 
\multicolumn{2}{c}{\TopI[\ if $A\sRew_\R^*\top$]$A$\DisplayProof
\hfill
    \AxiomC{$ A$}
    \BotE[\ if $A\sRew_\R^*\bot$]$ B$
    \DisplayProof\hfill
   \Tnd[\ if $A\sRew_\R^*B\vee(B\imp\bot)$]$A$
    \DisplayProof}
  \end{tabular}
\end{center}
\caption{Inference Rules of Natural Deduction Modulo $\rewsys R$.}\label{fig:natmod}
\renewcommand{\arraystretch}{1}
  \end{figure*}

The length of a proof is the number of inferences used in it. We will
denote by $\mathcal T\Nvdash{k}_\R P$ the fact that there exists a
proof of $ P$ of length at most $k$ using a finite subset of $\mathcal
T$ ($\mathcal T$ can be infinite) as assumptions. In the case where
$\R=\emptyset$, we are back to pure natural deduction, and we will use
$\mathcal T\Nvdash{k} P$. 

\begin{definition}[{Compatible presentation \citep[Definition
    1.4]{DHK-TPM-JAR}}] An axiomatic presentation $\Gamma$ of a theory
  is called \Def{compatible} with a rewrite system $\R$ if:
  \begin{enumerate}[$\bullet$]
  \item $P\sRew_\R^*Q$ implies $\Gamma\Nvdash{\,} P\Leftrightarrow Q$;
  \item for every proposition $P\in\Gamma$, we have $\Nvdash{\,}_\R P$.
  \end{enumerate}
\end{definition}

For instance, $B\imp A$ is compatible
with $A\ra A\vee B$: it possible to prove $A\Leftrightarrow A\vee
B$ assuming $B\imp A$ with the proof:
  \begin{prooftree}
    \Axiomm[i]$A$
    \OrI$A\vee B$
    \ImpI[(i)]$A\imp A\vee B$
    \Axiomm[ii]$A\vee B$
    \Axiomm[iii]$A$
    \Axiomm[iii]$B$ \AxiomC{$B\imp A$}
    \ImpE$A$
    \OrE[(iii)]$A$
    \ImpI[(ii)]$A\vee B\imp A$
    \AndI$A\Leftrightarrow A\vee B$
  \end{prooftree}
(other cases of equivalent propositions\ can be derived from it),
and reciprocally, $B\imp A$ has the following proof modulo $A\ra A\vee
B$:
\begin{prooftree}
  \Axiomm[i]$B$
  \OrI[$A~\rewrite~A\vee B$]$A$
  \ImpI[(i)]$B\imp A$
\end{prooftree}

Given a rewrite system, a compatible presentation always exists: a
proposition rewrite rule $A\rightarrow B$ (resp. a term rewrite rule
$l\rightarrow r$) corresponds to an axiom
$\forall \vect x.~A\Leftrightarrow B$ (resp. $\forall\vect x.~l=r$)
where $\vect x$ are the free variables of $A$ (resp. $l$).
One can
show that proving modulo a rewrite system is the same as proving using
a compatible presentation as axioms~\citep[Proposition 1.8]{DHK-TPM-JAR}.

Proof lengths in finite compatible presentations are essentially the
same:
\begin{proposition}\label{prop:fin_sim}
  Let $\Gamma_1$ and $\Gamma_2$ be two finite presentations compatible
  with the same rewrite system $\rewsys R$.  It is possible to
  translate a proof of length $n$ in $\Gamma_1$
  into a proof of length $\bigO(n)$ in $\Gamma_2$.
$$\Gamma_1\Nvdash{k}P~~\leadsto~~\Gamma_2\Nvdash{\bigO(k)}P$$
\end{proposition}
\begin{proof}
  We show that every axiom of $\Gamma_1$ can be translated into a proof of
  bounded depth in $\Gamma_2$. By definition of compatibility, for all
  $P\in\Gamma_1$, we have a proof $\Nvdash{\,}_\R P$. Then, whenever the
  congruence is used in that proof, we replace it by a cut with the
  corresponding proof in $\Gamma_2$ thanks to compatibility. For
  instance, if we have
  \begin{prooftree}
    \AxiomC{$\varpi$}
    \noLine
    \UnaryInfC{$ A$} \AllE[with $A\sRew_\R^*\forall x.~C$ and
    $B\sRew_\R^*\subst Ctx$]$ B$
  \end{prooftree} we know by compatibility that there exists proofs
  $\pi_1$ of $\Gamma_2\Nvdash{\,} A\Leftrightarrow \forall x.~C$ and
  $\pi_2$ of $\Gamma_2\Nvdash{\,} B\Leftrightarrow \subst Ctx$, so
  that we have
  \begin{prooftree}
\AxiomC{$\pi_2$}
    \noLine
    \UnaryInfC{$B\Leftrightarrow \subst Ctx$}
\AndE$\subst Ctx\imp B$
\AxiomC{$\pi_1$}
    \noLine
    \UnaryInfC{$A\Leftrightarrow \forall x.~C$}
    \AndE$A\imp\forall x.~C$
    \AxiomC{$\varpi$}
    \noLine
    \UnaryInfC{$ A$}
    \ImpE$\forall x.~C$
\AllE$ \subst Ctx$
\ImpE$B$
  \end{prooftree}
  Transforming all applications of the congruence in that way, we
  obtain a proof $\pi_P$ of $\Gamma_2\Nvdash{\,}P$. As $\Gamma_1$ is finite,
  there is a maximum $K$ on the length of such proofs, and a proof of
  length $n$ in $\Gamma_1$ can be transformed into a proof of length
  at most $K\times n$ in $\Gamma_2$ by replacing an axiom $P$ by its
  corresponding proof $\pi_P$.
\end{proof}
\begin{note}
  This proposition holds also if one considers only cut-free
  proofs. Indeed, even if the proof $\pi_P$ above contains cuts, it is
  possible to eliminate them to obtain a proof $\varpi_P$. (Indeed,
  $\pi_P$ is a proof in standard natural deduction.)\enspace The
  resulting proof may be much bigger, but we only do so for the finite
  number of $P$ in $\Gamma_1$. Therefore, there remains a constant $K'$
  bounding the length of such proofs $\varpi_P$, and replacing the
  axioms $P$ by the proofs $\varpi_P$ in a cut-free proof of size $n$ in
  $\Gamma_1$ leads to a cut-free proof in $\Gamma_2$ of size $K'\times n$.
\end{note}

\subsection{A Simple Proof-Length Speed-up}

Because part of the proofs are put into the congruence, it is quite
easy to get arbitrary proof-length speed-ups in deduction modulo, even
for very simple rewrite systems. 

Consider the proposition rewrite system $$\rewsys Add \df \left\{
  \begin{aligned}
    Add(O,y,y)&\rightarrow \top\\
    Add(s(x),y,s(z))&\rightarrow Add(x,y,z)
  \end{aligned}\right.\enspace.$$
It is easy to prove that the derivational complexity of $\rewsys Add$
is polynomially bounded. Furthermore, it is confluent, and
$\sRew_{\rewsys Add}^*$ is clearly decidable in polynomial
time. However, proving modulo $\rewsys Add$ leads to an arbitrary
proof-length speed-up compared to proving using a finite compatible
presentation.

  \begin{theorem}\label{theo:simple}
   There is an infinite family $\mathcal F$ of
   propositions such that for all finite axiomatic presentations $\Gamma$
   compatible with $\rewsys Add$, 
   \begin{enumerate}[\em(1)]
   \item for all $P\in\mathcal F$, $\Gamma\Nvdash{} P$
   \item for all $ P\in\mathcal F$, $\Nvdash{1\text{ step}}_{\rewsys Add} P$
   \item there is no fixed $k\in\mathbb N$ such that for all
     $ P\in\mathcal F$, ${\Gamma\Nvdash{k\text{ steps}} P}$.
   \end{enumerate}
  \end{theorem}
  \begin{proof}
Let $\underline n$ denote $\underbrace{s(\cdots s(}_{n\text{ times}}
O){\cdot}{\cdot}{\cdot})$ for $n\in\mathbb N$.
   Consider the following family of propositions $\left(Add(\underline
     i,\underline i,\underline{2i})\right)_{i\in\mathbb N}$. Clearly,
   (1) holds. Since $Add((\underline
     i,\underline i,\underline{2i})\rewrite_{\rewsys Add}^* Add(O, \underline
     i,\underline i)\rewrite_{\rewsys Add} \top$, we have the
     following proof modulo $\rewsys Add$:
     \begin{prooftree}
       \TopI[$Add((\underline i,\underline
       i,\underline{2i})\sRew_{\rewsys Add}^* \top$]$Add(\underline
       i,\underline i,\underline{2i})$ \end{prooftree} Hence, (2)
     holds.  Consider the presentation containing the two axioms
     $\forall x.~Add(x,O,x)$ and $\forall
     x~y~z.~Add(s(x),y,s(z))\Leftrightarrow Add(x,y,z)$. It is easy to
     prove that this finite presentation is compatible with $\rewsys
     Add$. To prove $Add(\underline i,\underline i,\underline{2i})$ in
     this presentation, we need to use the second axiom at least $i$
     times, so that the length of the proofs cannot be bounded by a
     constant. Now consider another finite presentation compatible
     with $\rewsys Add$, Proposition~\ref{prop:fin_sim} tells us that
     the length of the proofs cannot be bounded by a constant in that
     presentation neither.
  \end{proof}
  \begin{note}
    The theorem above is not true for infinite compatible presentations, since such
    presentations can contain $\mathcal F$.
  \end{note}
\section{Proof systems for \texorpdfstring{$i$\ith}{ith}-order arithmetic}\label{sec:arith}
In higher-order arithmetic, one wants to reason about natural numbers,
but also about properties of these numbers, and properties of these
properties, etc.  There are several way to present higher-order
arithmetic. One of them is to define it as a theory of higher-order
logic, that is, with the possibility to quantify over propositions. In
that setting, the induction schema can be expressed as an axiom
$\forall P^{\iota\rightarrow o}.~P(0)\wedge(\forall
\beta^{\iota}.~P(\beta)\imp P(s(\beta)))\imp\forall
\alpha^{\iota}.~P(\alpha)$. It is also possible to consider Girard's
System F as a system for second-order arithmetic. In this paper, we
use another presentation of higher-order arithmetic which is more
common when speaking about proof length,  and which consists of a
first-order theory presented by what is called a schematic system. The
idea is to use comprehension axioms to link each proposition $A$ to a
first-order object $\alpha$, which can be thought of as the set of
terms satisfying the proposition:
\begin{gather*}
\exists\alpha^{j+1}.~\forall\beta^j.~
\beta\in\alpha\Leftrightarrow A(\beta)
\qquad(\alpha\text{ is not free in }A(\beta))
\end{gather*}
There are therefore several layers of terms: the one in which live the
natural numbers (which corresponds to the sort 0 below), the one in
which live the sets of natural numbers (sort~1), the one for the set
of set of natural numbers, etc.  Then, to quantify over a proposition,
one has to quantify over its corresponding set. For instance, the
induction schema could be presented as $\forall s^1.~0\in
s\wedge(\forall \beta^0.~{\beta\in s}\imp {s(\beta)\in
  s})\imp\forall\alpha^0.~\alpha\in s$. Notwithstanding, we do not do
so in the following to have a definition of $i$-th order arithmetic
that works also when $i=1$.

\subsection{Schematic systems}

We recall here, using \citeauthor{buss94goedel}'~\citeyear{buss94goedel} terminology, what a
schematic system consists of. It is essentially an Hilbert-type (or
Frege) proof system, i.e.\ valid propositions\ are derived from a finite
number of axiom schemata using a finite number of inference
rules. Theorem \ref{theo:buss} is true on condition that proofs are
performed using a schematic system. \medskip

Given a many-sorted signature of first-order logic, we can
consider infinite sets of \Def{metavariables} $\alpha^i,\beta^i,\gamma^i,\dots$ for each sort
$i$ (which will be substituted by variables), of \Def{term variables}
$\tau^i$ for each sort $i$ (which will be substituted by terms) and
\Def{proposition variables} $A(\vect x)$ for each arity $[\vect i]$
(which will be substituted by propositions).

Metaterms are built like terms, except that they can contain
metavariables and term variables. Metapropositions\ are built like
propositions, except that they can contain proposition variables (which
play the same role as predicates) and metaterms, and that they can
bind metavariables.

A \Def{schematic system} is a finite set of inference rules, where an
inference rule is a triple of a finite set of metapropositions\ (the \Def{premises}), a
metapropositions\ (the \Def{conclusion}), and a set of side conditions of the forms
\emph{$\alpha^j$ is not free in $\Phi$} or \emph{$s$ is freely
  substitutable for $\alpha^j$ in $\Phi$} where $\Phi$ is a
metaproposition and $s$ a metaterm of sort $j$. It is denoted by 
\begin{prooftree}
\AxiomC{$\Phi_1$} \AxiomC{$\cdots$} \AxiomC{$\Phi_n$}
\RightLabel{$(R)$}
\TrinaryInfC{$\Psi$}
\end{prooftree}
An inference with an empty set of premises will be called an \Def{axiom
schema}. An axiom schema without metaproposition is an \Def{axiom}.

\subsection{\texorpdfstring{$i$\ith}{ith}-order arithmetic}

$i$\ith-order arithmetic ($Z_{i-1}$) is a many-sorted theory with
sorts $0,\ldots,i-1$ and the
signature $$\begin{array}{r@{~:~}l@{\qquad\qquad}r@{~:~}l@{\qquad\qquad}r@{~:~}l}
  0 & 0& + & [0; 0]\ra 0&
  = & [0;0]\\
  s & [0]\ra 0& \times & [0; 0]\ra 0& \in^j & [j; j+1]
\end{array}\enspace.$$

The schematic system we use here consists of the following inference
rules:

\noindent\textbf{$\mathbf{15+2\times i}$ axiom schemata of classical
  logic.} These axiom schemata, together with the inference rules
below, defines classical many-sorted first-order logic with sorts
$0,\ldots,i-1$. We take those used by~\citet[Chapter
5]{gentzen:untersuchungen} to prove the equivalence of his formalisms
with an Hilbert-type proof system:
\begin{gather}\label{eq:I}
\tag{I}A\imp A
\\
\label{eq:K}
\tag{K}A\imp B\imp A
\\\label{eq:Co}
\tag{W}(A\imp A\imp B) \imp A \imp B
\\
\label{eq:C}
\tag{C}(A\imp B\imp C) \imp B\imp A\imp C
\\
\label{eq:T}
\tag{B}(A\imp B) \imp (B \imp C) \imp A \imp C
\\\label{eq:AndLL}
\tag{Proj$_l$}(A\wedge B) \imp A
\\\label{eq:AndLR}
\tag{Proj$_r$}(A\wedge B) \imp B
\\\label{eq:AndR}
\tag{Pair}(A\imp B) \imp (A\imp C)\imp A \imp (B\wedge C)
\\\label{eq:OrRL}
\tag{Inj$_l$}A\imp (A\vee B)
\\\label{eq:OrRR}
\tag{Inj$_r$}B\imp (A\vee B)
\\\label{eq:OrL}
\tag{Case}(A\imp C) \imp (B\imp C) \imp (A\vee B) \imp C
\\\label{eq:BotR}
\tag{Contradiction}(A\imp B)\imp (A\imp B\imp\bot)\imp A\imp\bot
\\\label{eq:BotL}
\tag{EFSQ}(A\imp\bot)\imp A\imp B
\\
\tag{T}\top
\\\label{eq:AllL}
\tag{UI}(\forall \alpha^j.~A(\alpha^j)) \imp A(\tau^j)
\\\nonumber\left(\tau^j\text{ is freely substitutable for }\alpha^j\text{ in
}A(\alpha^j)\right)
\\\label{eq:ExR}
\tag{EI}A(\tau^j) \imp \exists \alpha^j.~A(\alpha^j)
\\\nonumber\left(\tau^j\text{ is freely substitutable for }\alpha^j\text{ in
}A(\alpha^j)\right)
\\\label{eq:tnd}
\tag{TND}A\vee (A\imp\bot)
\end{gather}

\noindent\textbf{$\mathbf{1+2\times i}$ inference rules of classical
  logic.} They are the only inference rules of our schematic system. Again, we take those used by
\citet{gentzen:untersuchungen}:
\begin{equation}\label{eq:mp}
\tag{MP}\text{\AxiomC{$A$} \AxiomC{$A\imp B$}
\BinaryInfC{$B$}
\DisplayProof}
\end{equation}
\begin{equation}\label{eq:gen}
\tag{Gen}\text{\AxiomC{$A\imp B(\beta^j)$}
\RightLabel{($\beta^j$ is not free in
  $A\imp\forall\alpha^j.~B(\alpha^j)$)}
\UnaryInfC{$A\imp\forall\alpha^j.~B(\alpha^j)$}
\DisplayProof}
\end{equation}
\begin{equation}\label{eq:part}
\tag{Part}\text{\AxiomC{$B(\beta^j)\imp A$}
\RightLabel{($\beta^j$ is not free in $(\exists\alpha^j.~B(\alpha^j))\imp A$)}
 \UnaryInfC{$(\exists\alpha^j.~B(\alpha^j))\imp A$}
\DisplayProof}
\end{equation}

\noindent\textbf{2 identity axiom schemata.} They define the particular relation $=$:
\begin{gather}\label{eq:refl}
\tag{Refl}\forall\alpha^0.~ \alpha^0 = \alpha^0
\\\tag{Leibniz}\label{eq:cong_p}
\forall\alpha^0\beta^0.~\alpha^0=\beta^0\imp A(\alpha^0)\imp A(\beta^0)
\end{gather}

\noindent\textbf{7 Robinson's axioms.} They are the axioms defining the
function symbols of arithmetic~\citep{tarski53undecidable}:
\begin{gather}\label{eq:s_inj}
\tag{$0\neq s$}\forall\alpha^0.~\neg~0=s(\alpha^0)
\\
\tag{Inj$_s$}\label{eq:s_inv}
\forall\alpha^0\beta^0.~s(\alpha^0)=s(\beta^0)\imp\alpha^0=\beta^0
\\\tag{Onto$_s$}\label{eq:s_surj}
\forall\alpha^0.~(\neg~\alpha^0=0)\imp\exists\beta^0.~\alpha^0=s(\beta^0)
\\
\tag{$+0$}\label{eq:plus_0}
\forall\alpha^0.~\alpha^0+0=\alpha^0
\\\tag{$+s$}\label{eq:plus_s}
\forall\alpha^0\beta^0.~\alpha^0+s(\beta^0)=s(\alpha^0+\beta^0)
\\\tag{$\times 0$}\label{eq:mult_0}
\forall\alpha^0.~\alpha^0\times 0=0
\\\tag{$\times s$}\label{eq:mult_s}
\forall\alpha^0\beta^0.~\alpha^0\times
s(\beta^0)=\alpha^0\times\beta^0 + \alpha^0
\end{gather}

\noindent\textbf{$\mathbf{i+1}$ induction and comprehension axiom schemata.} 
The induction schema is essential to have first-order arithmetic, and
not Robinson's arithmetic that is considerably weaker. It allows for
instance to prove $\forall \alpha^0.~s(\alpha^0)\neq\alpha^0$.
\begin{equation}\label{eq:ind}\tag{Ind}
  A(0)\imp\left(\forall\beta^0.~A(\beta^0)\imp
  A(s(\beta^0))\right)\imp \forall \alpha^0.~A(\alpha^0)
\end{equation}

The comprehension axiom schemata permits to introduce higher-order
objects up to order $i$.
For all $0\leq j<i-1$,
\begin{equation}\label{eq:comp}\tag{Comp$^j$}
\exists\alpha^{j+1}.~\forall\beta^j.~
\beta^j\in^j\alpha^{j+1}\Leftrightarrow A(\beta^j)
\qquad(\alpha^{j+1}\text{ is not free in }A(\beta^j))
\end{equation}

From this point on, we will denote by $Z_{i-1}\Hvdash{k} P$ the fact
that there exists a proof of $ P$ of length at most $k$ in this
schematic system, i.e. $P$ can be derived using at most $k$ instances
of these inference rules.  Abusing notations, we will write
$Z_{i-1}\Nvdash{k} P$ to say that there is a proof of $ P$ in natural
deduction of length at most $k$ using as assumptions a finite subset
of instances of the axiom schemata~(\ref{eq:refl}),
\eqref{eq:cong_p}, Robinson's axioms, \eqref{eq:ind} and (\ref{eq:comp}).

\subsection{Translations between schematic systems and natural
  deduction}\label{sec:trans}
\hyperlink{hl:theo:buss}{Buss' theorem} is true in schematic systems, but deduction modulo is
mostly studied in natural deduction or in the sequent calculus. It is
important to get bounded translations between these formalisms to show
that the speed-ups we will be considering are not artifacts of the
deductive system.

\subsubsection{From \texorpdfstring{$Z_i\Hvdash{}$}{schematic systems}
  to \texorpdfstring{$Z_i\Nvdash{\,}$}{natural deduction}}\label{sec:HtoN}
We want to translate a proof in the schematic system of $Z_i$ into a
proof in pure natural deduction using as assumptions instances of the
axiom schemata~(\ref{eq:refl}) to~(\ref{eq:comp}).

For the axiom schemata and inference rules of classical logic, we use
the same translation as Gentzen, for instance the axiom schema
(\ref{eq:C}) is translated into the natural deduction proof
\begin{prooftree}
  \Axiomm[ii]$B$
  \Axiomm[iii]$A$
  \Axiomm[i]$A\imp B\imp C$
  \ImpE$B\imp C$
  \ImpE$C$
  \ImpI[(iii)]$A\imp C$
  \ImpI[(ii)]$B\imp A\imp C$
  \ImpI[(i)]$(A\imp B\imp C)\imp B\imp A\imp C$
\end{prooftree}
and the inference rule (\ref{eq:part}) into 
\begin{prooftree}
  \Axiomm[i]$\exists\alpha^j.~B(\alpha^j)$
  \Axiomm[ii]$B(\beta^j)$
  \AxiomC{$B(\beta^j)\imp A$}
  \ImpE$A$
  \ExE[(ii)]$A$
  \ImpI[(i)]$\exists\alpha^j.~B(\alpha^j)\imp A$
\end{prooftree}
 (note that the side condition ensure that it is
possible to consider that what will be substituted for $\beta$ is free in
$A$ and the assumptions of the proof above $B(\beta^j)\imp A$).
All these inference rules have a translation whose length does
not depend on the propositions\ finally substituted in the proof.

In a schematic system proof, there is also a finite number of
instances of the axiom schemata for identity, Robinson's axioms and
induction and comprehension schemata. We keep these instances as
assumptions in natural deduction, so that we obtain a proof in
natural deduction using as assumptions a finite subset of instances of the
axiom schemata~(\ref{eq:refl}) to~(\ref{eq:comp}), and whose length
is linear compared to the schematic system proof:
\begin{proposition}\label{prop:HtoN}
  It is possible to translate a proof of length $n$ in the schematic
  system for $Z_i$ into a proof of length $\bigO(n)$ in (pure) natural
  deduction using assumptions in $Z_i$.
\medskip

{\hfill $Z_i\Hvdash{k} P~~\leadsto~~Z_i\Nvdash{\bigO(k)} P$\qed}

\end{proposition}

\subsubsection{From \texorpdfstring{$Z_i\Nvdash{\,}$}{natural
    deduction} to \texorpdfstring{$Z_i\Hvdash{}$}{schematic systems}}\label{sec:NtoH}
In this section, we consider a proof of $P$ in natural deduction,
using as assumption finite instances of (\ref{eq:refl})
to~(\ref{eq:comp}) in the language of $Z_i$. We translate it into a
proof in the schematic system for $Z_i$.

This is essentially a generalization of the translation from the
$\lambda$-calculus to combinatory logic
\citep[see][]{Curry-et-al-58}. We define mutually recursively two
functions by induction on the inference rules: T transforms a proof of
$ P$ in natural deduction using assumptions $\Gamma$ into a proof of $
P$ in the schematic system (\ref{eq:I}) to (\ref{eq:part}) plus
$\Gamma$. T$_A$ transform a proof of $ P$ in natural deduction using
assumptions $\Gamma, A$ into a proof of $A\imp P$ in the schematic
system consisting of the rules (\ref{eq:I}) to (\ref{eq:part}) and
the propositions of $\Gamma$ (seen as axioms).  The translation
can be found in the appendix.

It can be verified that this transformation is at most exponential in
the length of proofs. Due to \citet[Corollary~3.4]{cook79relative}, we
could have found, at least for the propositional part, a polynomial
translation. Nevertheless all we need in this paper is the fact that
the increase of the proof length in the translation is bounded.

\begin{proposition}\label{prop:NtoH}
  There exists some constant $K$ such that it is possible to translate a proof of length $n$ in the (pure) natural
  deduction using assumptions in $Z_i$ into a proof of length $\bigO(K^n)$ in the schematic
  system for $Z_i$.
$$Z_i\Nvdash{k} P~~\leadsto~~Z_i\Hvdash{\bigO(K^k)} P$$
\end{proposition}
\begin{proof}
  Let $K$ be the maximum number of steps that appear in addition of
  the recursive calls in the definition of T$_A$ (note that it does
  not depend on $A$). First, if a proof $\varpi$ does not contain
  $\imp$-i, $\vee$-e or $\exists$-e, then $|T_A(\varpi)|\leq
  K|\varpi|$. We prove this by induction on $\varpi$. Let us detail
  $\imp$-e only, using notations of the appendix, the other cases
  being similar: \def\taille#1{|#1|}
\begin{align*} \taille{\Tm[A]{\varpi}} &=
  \taille{\Tm[A]{\pi_1}}+\taille{\Tm[A]{\pi_2}}+ 7\\ &\leq
  K\taille{\pi_1}+K\taille{\pi_2} + K & \text{by induction hypothesis,
    and by definition of }K\\ &\leq
  K\left(\taille{\pi_1}+\taille{\pi_2}+1\right)\\ &\leq
  K\taille{\varpi}
\end{align*}
Now let us show that in all cases $|T_A(\varpi)|\leq
  K^{|\varpi|}$. This is also proved by induction on $\varpi$. We
  only detail the case of
  $\imp$-i. $\taille{\Tm[A]{\varpi}}=\taille{\Tm[A]{\Tm[B]{\pi}}}$. By
  induction hypothesis, $\taille{\Tm[B]\pi}\leq
  K^{|\pi|}$. Furthermore, $\Tm[B]{\pi}$ does contain neither
  $\imp$-i, $\vee$-e nor $\exists$-e, so that
  $\taille{\Tm[A]{\Tm[B]{\pi}}}\leq K\taille{\Tm[B]{\pi}} \leq K\times
  K^{|\pi|} = K^{|\pi|+1}=K^{|\varpi|}$. From this result, we can deduce
  the bound for T.
\end{proof}

\section{Higher-order arithmetic as a first-order theory modulo}\label{sec:main}
In this section, we want to express higher-order arithmetic as a
rewrite system, while preserving the length of proofs. We first encode
everything related to higher orders into a rewrite system, keeping
axioms concerned only with first order. Second, we show how to orient
the remaining axioms as rewrite rules, therefore obtaining a rewrite
system encoding higher-order arithmetic as a whole.

\subsection{Encoding higher orders using classes}\label{sec:HtoNRi}
\newcommand{\inc}{~\upepsilon~}%
\newcommand{\inlineinc}{\upepsilon}%
First, we want to toss away every axioms that include a higher-order
symbol by translating them into rewrite rules. We also want to keep a
finite number of axioms, and not for instance axiom schemata. Indeed,
first-order theorem provers generally cannot handle such schemata.
Therefore, we want to obtain a presentation of higher-order
arithmetic with a finite number of first-order-only axioms, resorting
to the congruence to get the higher orders again.

To do so, we first consider the theory consisting of the \emph{axioms}
in (\ref{eq:refl}) to~(\ref{eq:mult_s}), so without the axiom schemata
(\ref{eq:cong_p}), (\ref{eq:ind}) and (\ref{eq:comp}) that corresponds
to an infinite number of axioms. Those are replaced by three new
axioms (\ref{eq:cong_p_c}), (\ref{eq:ind_c}) and (\ref{eq:comp_c}).
To do so, we use the work of \citet{fk:classes06} which permits to
express first-order theories using a finite number of axioms. The idea
is to transform each metaproposition $A(\vect t)$ used in an axiom
schema into a proposition of the form $\langle\vect
t\rangle\inc\gamma$ where $\gamma$ is a variable that will be
instantiated by a term representing what
proposition is actually substituted for $A$. Such a term is called a
class, by reference to set theory where a class is a collection of
sets defined by some property they share.  Note that is long known
that using classes permits to have finite first-order axiomatization
\citep[see for instance][]{kleene52finite}, but Kirchner's work shows
how to handle the classes with a simple rewrite system of weak
explicit substitutions.

Following Kirchner's method, we add the new sorts $\ell$ for lists and
$c$ for classes, as well as the new function symbols and predicate
\begin{center}
$\begin{array}{r@{~:~}l}
    1^j & j\\
    S^j & [j]\ra j\\
    \cdot[\cdot]^j & [j; \ell]\ra j\\
  \end{array}\hfill
  \begin{array}{r@{~:~}l}nil & \ell\\
::^j & [j;\ell]\ra\ell\\
    \doteq & [0;0]\ra c\\
    \dot{\in}^j & [j;j+1]\ra c\\
  \end{array}\hfill
  \begin{array}{r@{~:~}l}\cup & [c;c]\ra c\\
    \cap & [c;c]\ra c\\
    \supset & [c;c]\ra c\\
  \end{array}\hfill
  \begin{array}{r@{~:~}l}\emptyset & c\\
    \mathcal P^j & [c]\ra c\\
    \mathcal C^j & [c]\ra c\\
    \inlineinc & [\ell;c]
  \end{array}\enspace.$
\end{center}
$\langle\vect\alpha\rangle$ will be syntactic sugar for
${\alpha_1::^{j_1}\cdots::\alpha_n::^{j_n}nil}$ for the appropriate
$j_m$. Note that we only need one sort of class, and not one for each
order, as we could have done. That way, all substitutions are done in
the same setting.  We change the axiom schemata (\ref{eq:cong_p}),
(\ref{eq:ind}) and (\ref{eq:comp}) into the following \emph{axioms}:
\begin{equation}\label{eq:cong_p_c}\tag{Leibniz$_{ax}$}
\forall\gamma^c.~\forall\alpha^0\beta^0.~\alpha^0=\beta^0\imp \langle\alpha^0\rangle\inc\gamma^c\imp\langle\beta^0\rangle\inc\gamma^c
\end{equation}
\begin{equation}\label{eq:ind_c}\tag{Ind$_{ax}$}
    \forall \gamma^c.\langle 0\rangle\inc\gamma^c\imp\left(\forall\beta^0.~\langle\beta^0\rangle\inc\gamma^c\imp
    \langle s(\beta^0)\rangle\inc\gamma^c\right)\imp \forall
    \alpha^0.~\langle\alpha^0\rangle\inc\gamma^c
  \end{equation}
For all $0\leq j<i$,
\begin{equation}\label{eq:comp_c}\tag{Comp$^j_{ax}$}
\forall\gamma^c.~\exists\alpha^{j+1}.~\forall\beta^j.~
\beta^j\in^j\alpha^{j+1}\Leftrightarrow \langle\beta^j\rangle\inc\gamma^c
\end{equation}

We also need weak-substitution axioms which permit to decode the classes
\citep[see][Definition 4]{fk:classes06}.
\begin{align}\label{eq:ws_begin}
\tag{WS$_{nil}$}\forall \alpha^j.~~ \alpha^j[nil]^j&= \alpha^j\\
\tag{WS$_{1^j}$}\forall \alpha^j.~\forall l^\ell.~~ 1^j[\alpha^j::^jl^\ell]^j&= \alpha^j\\
\tag{WS$_{S^j}$}\forall \alpha^j.~\forall \beta^k.~\forall l^\ell.~~S^j(\alpha^j)[\beta^k::^kl^\ell]^j&= \alpha^j[l^\ell]^j\\
\tag{WS$_{s}$}\forall \alpha^0.~\forall l^\ell.~~s(\alpha^0)[l^\ell]^0&= s(\alpha^0[l^\ell]^0)\\
\tag{WS$_{+}$}\forall \alpha^0.~\forall \beta^0.~\forall l^\ell.~~(\alpha^0+\beta^0)[l^\ell]^0&= \alpha^0[l^\ell]^0+\beta^0[l^\ell]^0\\
\tag{WS$_{\times}$}\forall \alpha^0.~\forall \beta^0.~\forall l^\ell.~~(\alpha^0\times\beta^0)[l^\ell]^0&= \alpha^0[l^\ell]^0\times\beta^0[l^\ell]^0\\
\tag{WS$_{=}$}\forall \alpha^0.~\forall \beta^0.~\forall l^\ell.~~l^\ell\inc\doteq(\alpha^0,\beta^0)&\Leftrightarrow \alpha^0[l^\ell]^0 = \beta^0[l^\ell]^0\\
\tag{WS$_{\in^j}$}\forall \alpha^j.~\forall \beta^{j+1}.~\forall l^\ell.~~l^\ell\inc{\dot{\in}}^j(\alpha^j,\beta^{j+1})&\Leftrightarrow \alpha^j[l^\ell]^j \in^j \beta^{j+1}[l^\ell]^{j+1}\\
\tag{WS$_{\vee}$}\forall \alpha^c.~\forall \beta^c.~\forall l^\ell.~~l^\ell\inc \alpha^c\cup \beta^c&\Leftrightarrow l^\ell\inc \alpha^c\vee l^\ell\inc \beta^c\\
\tag{WS$_{\wedge}$}\forall \alpha^c.~\forall \beta^c.~\forall l^\ell.~~l^\ell\inc \alpha^c\cap \beta^c&\Leftrightarrow l^\ell\inc \alpha^c\wedge l^\ell\inc \beta^c\\
\tag{WS$_{\imp}$}\forall \alpha^c.~\forall \beta^c.~\forall l^\ell.~~l^\ell\inc \alpha^c\supset \beta^c&\Leftrightarrow l^\ell\inc \alpha^c\imp l^\ell\inc \beta^c\\
\tag{WS$_{\bot}$}\forall l^\ell.~~l^\ell\inc \emptyset&\Leftrightarrow \bot\\
\tag{WS$_{\exists^j}$}\forall \alpha^c.~\forall l^\ell.~~(l^\ell\inc \mathcal
P^j(\alpha^c)&\Leftrightarrow \exists \beta^j.~\beta^j::^jl^\ell\inc
\alpha^c)\\
\label{eq:ws_end}
\tag{WS$_{\forall^j}$}\forall \alpha^c.~\forall l^\ell.~~(l^\ell\inc \mathcal C^j(\alpha^c)&\Leftrightarrow \forall \beta^j.~\beta^j::^jl^\ell\inc \alpha^c)
\end{align}

\begin{definition}
  The axiomatic presentation $Z_i^{\textup{ws}}$ consists of the axioms
  (\ref{eq:refl}), Robinson's axioms, (\ref{eq:cong_p_c}),
  (\ref{eq:ind_c}), (\ref{eq:comp_c}) and all (WS) axioms.
\end{definition}
In other words, $Z_i^{\textup{ws}}$ is the finite axiomatic
presentation obtained by applying \citeauthor{fk:classes06}'s ideas to $Z_i$.

\begin{proposition}
The theory $Z_i^{\textup{ws}}$ is a conservative extension of $Z_i$.
\end{proposition}
\begin{proof}  This is the Proposition~4 of \citet{fk:classes06}.
\end{proof}

Now, we use skolemization to transform (\ref{eq:comp_c})
\citep[see][Section 3.4]{dalen89logic}. We add new function symbols
$comp^j:[c]\rightarrow j$ for all $0<j\leq i$. We then consider the
skolemized version of (\ref{eq:comp_c}):
\begin{equation}\label{eq:comp_c_s}\tag{Comp$^j_{sk}$}
\forall\gamma^c.~\forall\beta^j.~
\beta^j\in^jcomp^{j+1}(\gamma^c)\Leftrightarrow \langle\beta^j\rangle\inc\gamma^c
\end{equation}

\begin{definition}
  The axiomatic presentation $Z_i^{\textup{sk}}$ consists of the axioms
  (\ref{eq:refl}), Robinson's axioms, (\ref{eq:cong_p_c}),
  (\ref{eq:ind_c}), (\ref{eq:comp_c_s}) and all (WS) axioms.
\end{definition}
In other words, $Z_i^{\textup{sk}}$ is the presentation obtained by
skolemizing axiom \eqref{eq:comp_c} in $Z_i^{\textup{ws}}$.

\begin{proposition}
The theory $Z_i^{\textup{sk}}$ is a conservative extension of $Z_i^{\textup{ws}}$.
\end{proposition}
\begin{proof}According to \citet[Corollary 3.4.5]{dalen89logic},
  $Z_i^{\textup{sk}}\cup\{(\text{\ref{eq:comp_c}})\}$ is a conservative
  extension of $Z_i^{\textup{ws}}$. But (\ref{eq:comp_c}) can be
  proved in $Z_i^{\textup{sk}}$ so that we can drop it.
\end{proof}

We can then transform each axiom where a higher-order function symbol
or predicate appears, as well as each axiom decoding classes, into a
rewrite rule, and work modulo the resulting rewrite system. We denote
by $\HO_i$ the rewrite system defined in Figure~\ref{fig:hoi}.
\begin{figure}
$$\begin{array}{c@{~~~~~~~~~~~}c}
  \begin{array}{r@{~\ra~}l}t[nil]^j& t\\
    1^j[t::^jl]^j& t\\
    S^j(n)[t::^kl]^j& n[l]^j\\
    s(n)[l]^0& s(n[l]^0)\\
    (t_1+t_2)[l]^0& t_1[l]^0+t_2[l]^0\\
    (t_1\times t_2)[l]^0& t_1[l]^0\times t_2[l]^0\\
    l\inc\doteq(t_1,t_2)& t_1[l]^0 = t_2[l]^0\\
  \end{array}&
  \begin{array}{r@{~\ra~}l}
    l\inc{\dot{\in}}^{j'}(t_1,t_2)& t_1[l]^{j'} \in^{j'} t_2[l]^{j'+1}\\
    l\inc A\cup B& l\inc A\vee l\inc B\\
    l\inc A\cap B& l\inc A\wedge l\inc B\\
    l\inc A\supset B& l\inc A\imp l\inc B\\
    l\inc \emptyset& \bot\\
    l\inc \mathcal P^j(A)& \exists x.~x::^jl\inc A\\
    l\inc \mathcal C^j(A)& \forall x.~x::^jl\inc A
  \end{array}\\
  \multicolumn{2}{c}{x\in^{j'}comp^{j'+1}(A)\ra x::^{j'}nil\inc A}
\end{array}
$$
for all $0\leq j \leq i$, $0\leq k \leq i$ and $0\leq j' < i$.

\caption{Rewrite rules of $\HO_i$}\label{fig:hoi}
\end{figure}
This rewrite system has the following properties:
\begin{enumerate}[$\bullet$]
\item It is finite (for a given $i$).

\item It is terminating in a polynomial number of steps (Proposition~\ref{prop:polyHOi}).  

\item It is confluent: it terminates and it is locally confluent,
  since the only critical pairs,
of the form $f(\vect
t)\mathop{\longleftarrow}\limits_{\HO_i} f(\vect
t)[nil]\rewrite_{\HO_i} f(t_1[nil],\ldots,t_n[nil])$ where  ${f\in\{+;\times;s\}}$, are
easily joinable. 

\item It is left-linear, i.e. variables appears only once on the
  left-hand side of each rule.
\end{enumerate}

Before showing that $\HO_i$ has a polynomially bounded derivational
complexity, let us first see how $\HO_i$ works and, in particular, how
it can be used to encode propositions as terms. Proposition 2 of
\citet{fk:classes06} states that, for all propositions $P$ of the
language of $i$\ith-order arithmetic, and for all finite lists of
variables $\alpha_1^{j_1},\ldots,\alpha_n^{j_n}$, it is possible to
prove constructively $$\exists \beta^c.~\forall
\alpha_1^{j_1}\cdots\alpha_n^{j_n}.~\langle\alpha_1^{j_1},\ldots,\alpha_n^{j_n}\rangle\inc
\beta^c\Leftrightarrow P\enspace.$$ Hence, the proof of this
proposition shows us how to construct the witness for $\beta^c$. We
will denote it by $E_P^{\alpha_1^{j_1},\ldots,\alpha_n^{j_n}}$, and it
is therefore defined as:
\begin{align*}
  ||\alpha^j||^{\emptyset} &\df\alpha^j\\
  ||\alpha_1^{j_1}||^{\alpha_1^{j_1},\ldots,\alpha_n^{j_n}} &\df
  1^{j_1}\\
  ||\alpha^j||^{\alpha_1^{j_1},\ldots,\alpha_n^{j_n}} & \df S^j(||\alpha^j||^{\alpha_2^{j_2},\ldots,\alpha_n^{j_n}}) &&\text{if } \alpha^j\neq \alpha_1^{j_1}\\
  || 0||^{\emptyset}&\df  0\\
  || 0||^{\alpha_1^{j_1},\ldots,\alpha_n^{j_n}} &\df S^0(|| 0||^{\alpha_2^{j_2},\ldots,\alpha_n^{j_n}})\\
  ||s(t)||^{\alpha_1^{j_1},\ldots,\alpha_n^{j_n}} &\df s(||t||^{\alpha_1^{j_1},\ldots,\alpha_n^{j_n}})\\
  ||t_1+t_2||^{\alpha_1^{j_1},\ldots,\alpha_n^{j_n}} & \df ||t_1||^{\alpha_1^{j_1},\ldots,\alpha_n^{j_n}}+||t_2||^{\alpha_1^{j_1},\ldots,\alpha_n^{j_n}}\\
 ||t_1\times t_2||^{\alpha_1^{j_1},\ldots,\alpha_n^{j_n}} & \df ||t_1||^{\alpha_1^{j_1},\ldots,\alpha_n^{j_n}}\times||t_2||^{\alpha_1^{j_1},\ldots,\alpha_n^{j_n}}\\
  E_{t_1=t_2}^{\alpha_1^{j_1},\ldots,\alpha_n^{j_n}} & \df \doteq(||t_1||^{\alpha_1^{j_1},\ldots,\alpha_n^{j_n}}, ||t_2||^{\alpha_1^{j_1},\ldots,\alpha_n^{j_n}})\\
  E_{t_1\in^{j} t_2}^{\alpha_1^{j_1},\ldots,\alpha_n^{j_n}} & \df \dot\in^{j}(||t_1||^{\alpha_1^{j_1},\ldots,\alpha_n^{j_n}}, ||t_2||^{\alpha_1^{j_1},\ldots,\alpha_n^{j_n}})\\
  E_{\bot}^{\alpha_1^{j_1},\ldots,\alpha_n^{j_n}} &\df \emptyset \\
  E_{P\imp Q}^{\alpha_1^{j_1},\ldots,\alpha_n^{j_n}} & \df E_P^{\alpha_1^{j_1},\ldots,\alpha_n^{j_n}}\supset E_q^{\alpha_1^{j_1},\ldots,\alpha_n^{j_n}}\\
  E_{P\wedge Q}^{\alpha_1^{j_1},\ldots,\alpha_n^{j_n}} & \df E_P^{\alpha_1^{j_1},\ldots,\alpha_n^{j_n}}\cap E_q^{\alpha_1^{j_1},\ldots,\alpha_n^{j_n}}\\
  E_{P\vee Q}^{\alpha_1^{j_1},\ldots,\alpha_n^{j_n}} & \df E_P^{\alpha_1^{j_1},\ldots,\alpha_n^{j_n}}\cup E_q^{\alpha_1^{j_1},\ldots,\alpha_n^{j_n}}\\
  E_{\forall \alpha^j.~P}^{\alpha_1^{j_1},\ldots,\alpha_n^{j_n}} & \df \mathcal C^j(E_P^{\alpha^j,\alpha_1^{j_1},\ldots,\alpha_n^{j_n}}) &&\text{if }
  \alpha^j\not\in\{\alpha_1^{j_1},\ldots,\alpha_n^{j_n}\}\\
  E_{\exists \alpha^j.~P}^{\alpha_1^{j_1},\ldots,\alpha_n^{j_n}} & \df \mathcal P^j(E_P^{\alpha^j,\alpha_1^{j_1},\ldots,\alpha_n^{j_n}}) &&\text{if }
  \alpha^j\not\in\{\alpha_1^{j_1},\ldots,\alpha_n^{j_n}\}
\end{align*}
Then, one can prove that $\langle \vect t\rangle\inc
E_P^{\vect \alpha}\rewrite^*\{t_1/\alpha_1,\ldots, t_n/\alpha_n\}P$.

 For instance,
consider the proposition $P\df x = 0\vee\exists y.~x \in^0y$. Then $E_P^x$
equals $\doteq(1^0,S^0(0))~\cup~\mathcal
P^1\left(\dot{\in}^0(S^0(1^0),1^1)\right)$ and $\langle t\rangle\inc E_P^x$
can be rewritten to $t = 0\vee\exists x^1.~t\in^0x^1$:
\allowdisplaybreaks[0]
\begin{align*}
  \langle t\rangle\inc E_P^x&~\rewrite~ \langle t\rangle\inc
  \doteq(1^0,S^0(0))~\vee~ \langle t\rangle\inc \mathcal
P^1\left(\dot{\in}^0(S^0(1^0),1^1)\right)\\
& ~\rewrite^2~ \left(1^0[t::^0nil] = S^0(0)[t::^0nil]\right) \vee \exists x^1.~\langle
x^1,t\rangle\inc\dot{\in}^0(S^0(1^0),1^1)\\
& ~\rewrite^3~ \left(t = 0[nil]\right) \vee \exists x^1.~S^0(1^0)[x^1::^1t::^0::nil]
\in^0 1^1[x^1::^1t::^0::nil]\\
&~\rewrite^3~ t = 0 \vee \exists x^1.~1^0[t::^0::nil] \in^0 x^1\\
&~\rewrite~ t = 0 \vee \exists x^1.~t\in^0 x^1
\end{align*}
\allowdisplaybreaks[4]

\begin{proposition}\label{prop:polyHOi}
 The derivational complexity of  $\HO_i$ is polynomially bounded.
\end{proposition}
\begin{proof}
  Let us note $\WS_i$ the system $\HO_i$ without the last
  rule. $\WS_i$ is computing the application of a
  substitution to (the encoding of) a proposition. It cannot be
  applied more than the size of the right-hand side of $\inlineinc$
  and the left-hand side of $\cdot[\cdot]$ (by simple induction on the
  derivation). Therefore, the derivational complexity of $\WS_i$ is linear.  Now, note that a substitution is blocked by all
  $comp^j$, i.e.\ $comp^j(t)[l]$ cannot be reduced if neither $t$ nor
  $l$ can.  Therefore, the last rule of $\HO_i$ can only be applied to
  the outermost $comp^j$s: due to the sort constraints, $\in^j$ cannot
  appear inside a $comp^{j'}$, and if a $\dot\in^j$ is transformed
  into a $\in^j$ by the rule $l\inc\dot\in^j(t_1,t_2)\rightarrow t_1[l]^j \in^j
  t_2[l]^{j+1}$, the substitution applied to $t_2$ blocks 
  $comp^{j+1}$ if it is its function symbol.  Applying $\WS_i$ can duplicate the
  initially outermost $comp^j$s, but not more than the total number of
  $1^{j'}$ in the initial term. Once the last rule of $\HO_i$ is
  applied to all these copies of the outermost $comp^j$s, only $\WS_i$ can be applied. Therefore, the derivational complexity of
  $\HO_i$ is polynomially bounded.
\end{proof}

The axiom schemata (\ref{eq:cong_p}), (\ref{eq:ind}) and
(\ref{eq:comp}) can be replaced by the proofs in
Figure~\ref{fig:transc}.  Note that the replacement for
(\ref{eq:comp}) does not need extra axioms, because all is done in the
congruence.

\begin{figure*}
    \def\fCenter{\vdash_{\R_i}}
    \begin{prooftree}
      \Axiomm[\ref{eq:cong_p_c}]$\forall\gamma^c.~\forall\alpha^0\beta^0.~\alpha^0=\beta^0\imp \langle\alpha^0\rangle\inc\gamma^c\imp\langle\beta^0\rangle\inc\gamma^c$
      \AllE$\forall\alpha^0\beta^0.~\alpha^0=\beta^0\imp
      A(\alpha^0)\imp A(\beta^0)$
    \end{prooftree}
{\small\hfill(because $\langle\alpha^0\rangle\inc E_{A(x)}^{x}\imp\langle\beta^0\rangle\inc E_{A(x)}^{x}\rewrite^*A(\alpha^0)\imp A(\beta^0)$)}
\bigskip

    \begin{prooftree}
      \Axiomm[\ref{eq:ind_c}]$\forall \gamma^c.\langle 0\rangle\inc\gamma^c\imp\left(\forall\beta^0.~\langle\beta^0\rangle\inc\gamma^c\imp
      \langle s(\beta^0)\rangle\inc\gamma^c\right)\imp \forall \alpha^0.~\langle\alpha^0\rangle\inc\gamma^c$
      \AllE$A(0)\imp\left(\forall\beta^0.~A(\beta^0)\imp
      A(s(\beta^0))\right)\imp \forall \alpha^0.~A(\alpha^0)$
    \end{prooftree}
{\hfill\small(because for all $t$, $\langle t\rangle\inc E_{A(x)}^x\rewrite^*A(t)$)}
\medskip

    \small  \begin{prooftree}
      \Axiomm[i]$\beta^j\in^j comp^{j+1}(E_{A(x)}^x)$
      \ImpI[(i)]$\beta^j\in^j comp^{j+1}(E_{A(x)}^x)\imp\beta^j\in^j comp^{j+1}(E_{A(x)}^x)$
      \Axiomm[ii]$\beta^j\in^j comp^{j+1}(E_{A(x)}^x)$
      \ImpI[(ii)]$\begin{array}[b]{c}\beta^j\in^j comp^{j+1}(E_{A(x)}^x)\imp\beta^j\in^j
        comp^{j+1}(E_{A(x)}^x)\\\hfill\vdots~~\\\hfill\vdots~~\end{array}$
      \insertBetweenHyps{\kern-4cm}
      \kernHyps{1.5cm}    \AndI$\beta^j\in^j comp^{j+1}(E_{A(x)}^x)\Leftrightarrow\beta^j\in^j
      comp^{j+1}(E_{A(x)}^x)$

      \AllI$\forall\beta^j.~\beta^j\in^j comp^{j+1}(E_{A(x)}^x)\Leftrightarrow\beta^j\in^j
      comp^{j+1}(E_{A(x)}^x)$
      \ExI[$
        \begin{array}{@{}l@{}}
          \beta^j\in^jcomp^{j+1}(E_{A(x)}^x)\\\rewrite\langle\beta^j\rangle\inc
          E_{A(x)}^x\rewrite^*A(\beta^j)
        \end{array}
        $]$\exists\alpha^{j+1}.~\forall\beta^j.~
      \beta^j\in^j\alpha^{j+1}\Leftrightarrow A(\beta^j)$
    \end{prooftree}

  \caption{Translations of the axiom schemata (\ref{eq:cong_p}), (\ref{eq:ind}) and
(\ref{eq:comp}).}\label{fig:transc} 
\end{figure*}

\begin{definition}
  The finite axiomatic presentation $FZ$  consists of (\ref{eq:refl}), Robinson's
  axioms, (\ref{eq:cong_p_c})
  and (\ref{eq:ind_c}).
\end{definition}
\begin{note}
  All axioms of $FZ$ are in the language of $Z_0$ plus the language of
  \citeauthor{fk:classes06}'s classes.
\end{note}
$FZ$ can be seen as the first-order core of higher-order arithmetic,
whereas $\HO_i$  puts everything related to higher orders on.

 A proof $\pi$ of $ P$ in the
schematic system for $Z_i$ can be translated into a proof of $P$ in
natural deduction modulo $\HO_i$ using assumptions in $FZ$ whose
length is linear compared to the length of $\pi$.

\begin{proposition}\label{prop:HtoNRi}
  It is possible to translate a proof of length $n$ in the schematic
  system for $Z_i$ into a proof of length $\bigO(n)$ in the natural
  deduction modulo $\HO_i$ using assumptions in
  $FZ$.
$$Z_i\Hvdash{k}P~~\leadsto~~FZ\Nvdash{\bigO(k)}_{\HO_i}P$$
\end{proposition}
\begin{proof}
  Inference rules for classical first-order logic are translated as in
  Proposition~\ref{prop:HtoN}.  Instances of axiom schemata in the
  proof in $Z_i$ are replaced by the proofs in
  Figure~\ref{fig:transc}. The important point is that the length of
  these proofs does not depend on the particular instance that is
  considered.
\end{proof}
This result can also be stated entirely in natural deduction
\begin{theorem}\label{theo:s3}
  For all $i\geq 0$, there exists a finite confluent rewrite system
  with polynomially bounded derivational complexity $\HO_i$ such that
  for all propositions\ $ P$, if ${Z_i\Nvdash{k} P}$ then
  $FZ\Nvdash{\bigO(k)}_{\HO_i} P$.
  \end{theorem}
\begin{proof}
 We replace the instance of the axiom schemata (\ref{eq:cong_p}),
 (\ref{eq:ind}) and (\ref{eq:comp}) by proofs using the axioms
 (\ref{eq:cong_p_c}) and (\ref{eq:ind_c})  as
 indicated in Figure~\ref{fig:transc}. Here again, their length does
 not depend on the instance.
\end{proof}

\subsection{Higher-order arithmetic as purely computational theory} 
In this section, we define higher-order arithmetic entirely as a
rewrite system, modulo which inference rules are applied, without
resorting to any axiom. This is in line with the work of \citet{arith}
who express first-order arithmetic as a theory modulo. The idea is to
combine their work with the rewrite system of the previous section, to get a
characterization of higher-order arithmetic. Notwithstanding, we will
look carefully at the length of proofs in the translations.

\citet{arith} use the following method to introduce the induction
schema for first-order arithmetic: they add a new predicate $N$ of
arity $[0]$ which essentially states that an element is a natural
number, and thus can be used in the induction schema. $N(n)$ can
therefore be rewritten to $\forall p.~0\in p \imp\left(\forall y.~
  N(y)\imp {y\in p}\imp s(y)\in p\right)\imp n\in p$. Then, function
symbols $f_P^{x,\vect y}$ for each proposition $P$ of first-order
arithmetic with free variables $x,\vect y$ are added, as well as
rewrite rules $x\in f_P^{x,\vect y}(\vect y)\ra P$.  To prove a
proposition using induction, we need to know that the variables used in
the proof are natural numbers, hence quantifiers are relativized with
the predicate $N$ (i.e.\ $\forall x.~ P$ becomes $\forall x.~N(x)\imp
P$, and $\exists x.~P$ becomes $\exists x.~N(x)\wedge P$). Using this,
it is proved \citep[Proposition 13]{arith} that we obtain a
conservative extension of first-order arithmetic.  Nevertheless, the
length of the proofs is not preserved by the relativization. Indeed,
to translate a proof whose last step is
\begin{prooftree}
  \AxiomC{$\pi$}
  \noLine\UnaryInfC{$\forall x.~P$}
  \AllE$\subst Ptx$
\end{prooftree}
we have to transform it into a proof 
\begin{prooftree}
\AxiomC{$\varpi$}
\noLine\UnaryInfC{$N(t)$}
  \AxiomC{$|\pi|$}
  \noLine\UnaryInfC{$\forall x.~N(x)\imp P$}
\AllE$N(t)\imp\subst Ptx$
\ImpE$\subst Ptx$
\end{prooftree}
The problem is that the length of the proof $\varpi$ depends on the
size of $t$. In fact, it can be proved that there is an
arbitrary proof-length speed-up between the axiomatic presentation of
first-order arithmetic and the presentation of \citeauthor{arith}:
$\exists\alpha^0.~\alpha^0=\underline n$ can be proved in at most 7
steps in first-order arithmetic, whereas it needs a proof whose length
is linear in $n$ in the system of \citeauthor{arith}.

Hence, we use a different approach. Starting from $FZ$ modulo $\HO_i$,
it remains to orient the axioms of $FZ$ into rewrite rules. Axioms
\eqref{eq:plus_0} to \eqref{eq:mult_s} can be easily oriented. To
orient \eqref{eq:refl} and \eqref{eq:cong_p_c}, we use the axiom 
\begin{equation}
  \label{eq:eq_def}
  \tag{=$_{def}$} \forall \alpha^0~\beta^0.~\alpha^0=\beta^0
  \Leftrightarrow (\forall \gamma^c.~\langle
  \alpha^0\rangle\inc\gamma^c \imp\langle
  \beta^0\rangle\inc\gamma^c )
\end{equation}
which is equivalent to their conjunction. \eqref{eq:s_surj} is
redundant if the induction principle is present, so it can be dropped.
To encode \eqref{eq:s_inj} and \eqref{eq:s_inv}, we use the same
technique as \citet{arith}: we introduce a new function symbol $pred:
[0]\rightarrow 0$
and a new predicate $Null:[0]$, as well as new axioms defining them:
 \begin{gather}
\tag{$pred_0$}   \mathit{pred}(0) = 0\label{pred_0}\\
\tag{$pred_s$}   \forall \alpha^0.~\mathit{pred}(s(\alpha^0)) = \alpha^0\label{pred_s}\\
\tag{$Null_0$}   \mathit{Null}(0)\label{Null_0}\\
\tag{$Null_s$}   \forall \alpha^0.~\neg\mathit{Null}(s(\alpha^0))\label{Null_s}
 \end{gather}
which can be easily oriented.
It remains to orient the induction principle \eqref{eq:ind_c}. 
The most problematic part is that this axiom is the universal closure
of an implication, whereas proposition rewrite rules are compatible
with universal closures of logical equivalences where one of the side
of the equivalence is an atomic proposition. We use the fact that
$B\imp A$ is intuitionistically equivalent to $A\Leftrightarrow A\vee B$, so
that  \eqref{eq:ind_c} is equivalent to 
\begin{equation}
  \label{eq:ind_or}
  \tag{Ind$_{mod}$} \forall \alpha^0~\gamma^c.~\langle \alpha^0\rangle
  \inc \gamma^c \Leftrightarrow \left(\langle \alpha^0\rangle
  \inc \gamma^c \vee \left(\langle 0\rangle
  \inc \gamma^c \wedge (\forall \beta^0.~\langle \beta^0\rangle
  \inc \gamma^c \imp \langle s(\beta^0)\rangle
  \inc \gamma^c )\right)\right)
\end{equation}

If we do not use (\ref{eq:tnd}) as axiom (i.e.\ if we work
 in intuitionistic logic), we therefore obtain a formulation of
 higher-order Heyting arithmetic through the rewrite system
 $\HHA_i$ defined in Figure~\ref{fig:hha}.
 \begin{figure}
\shoveleft Arithmetic rules:
$$\begin{array}{r@{~\ra~}l}
  pred(0) & 0\\
  pred(s(x)) & x\\
  0+y & y\\
  s(x)+y & s(x+y)\\
\end{array}\hspace{2cm}
\begin{array}{r@{~\ra~}l}
  0\times y & y\\
  s(x)\times y & x\times y + y\\
  Null(0) & \top\\
  Null(s(x)) & \bot\\
\end{array}$$

\shoveleft Axiom schemata:
\begin{align*}
  x=y & \begin{aligned}\rightarrow\forall z^c.~\langle x\rangle\inc
    z\imp\langle y\rangle\inc z&\qquad& x\in^j comp^{j+1}(y)
    &\rightarrow x::^j nil\inc y
  \end{aligned}\\
  x::^0 nil\inc p &\rightarrow \langle x\rangle\inc p\vee(\langle
  0\rangle\inc p\wedge \forall y.~ \langle y\rangle\inc p\imp \langle
  s(y)\rangle\inc p)
\end{align*}

\shoveleft Substitutions and classes: $\WS_i$ + 
$$
\begin{array}{r@{~\ra~}l@{\qquad\qquad}r@{~\ra~}l}
  pred(n)[l]^0& pred(n[l]^0)&
  \ell\inc\dot{Null}(t) & Null(t[\ell]^0) 
\end{array}$$

\caption{Rewrite rules of $\HHA_i$}
\label{fig:hha}
\end{figure}
With this rewrite system, we can linearly simulate higher-order arithmetic in
deduction modulo:
\begin{theorem}\label{theo:hhamod}
For all $i$ there exists a finite rewrite system $\HHA_i$ such that for
all propositions $P$ in the language of $Z_i$, if $Z_i\Nvdash{k\text{
    steps}}P$ then $\Nvdash{\bigO(k)\text{ steps}}_{\HHA_i}P$.
\end{theorem}
\proof
It is sufficient to prove that all instances of the axiom schemata of
$Z_i$ can be proved in a bounded number of steps that does not depend
on the particular instance.

(\ref{eq:refl}) can be proved by
\begin{prooftree}
  \Axiomm[i]$\langle\alpha^0\rangle\inc p^c$
  \ImpI[(i)]$\langle\alpha^0\rangle\inc p^c\imp\langle\alpha^0\rangle\inc p^c$
  \AllI$\forall p^c.~\langle\alpha^0\rangle\inc p^c\imp\langle\alpha^0\rangle\inc p^c$
  \AllI[$\alpha^0=\alpha^0\rewrite\forall p^c.~\langle\alpha^0\rangle\inc
    p^c\imp\langle\alpha^0\rangle\inc
    p^c$]$\forall\alpha^0.~\alpha^0=\alpha^0$
\end{prooftree}

(\ref{eq:cong_p_c}) can be proved by
\begin{prooftree}
\Axiomm[i]$\langle\alpha^0\rangle\inc\gamma^c\imp\langle\beta^0\rangle\inc\gamma^c$
\ImpI[(i), $\alpha^0=\beta^0\rewrite
\langle\alpha^0\rangle\inc\gamma^c\imp\langle\beta^0\rangle\inc\gamma^c$]$\alpha^0=\beta^0\imp
\langle\alpha^0\rangle\inc\gamma^c\imp\langle\beta^0\rangle\inc\gamma^c$ 
\AllI[$3\times$]$\forall\gamma^c.~\forall\alpha^0\beta^0.~\alpha^0=\beta^0\imp
\langle\alpha^0\rangle\inc\gamma^c\imp\langle\beta^0\rangle\inc\gamma^c$ 
\end{prooftree}

(\ref{eq:s_inj}) is proved using
  $x=y \ra \forall z^c.~\langle x\rangle\inc z\imp\langle
y\rangle\inc z$:
\begin{prooftree}
 \Axiomm[i]$0=s(\alpha^0)$
 \AllE$\langle 0\rangle\inc \dot{Null}(1^0)\imp \langle
 s(\alpha^0)\rangle\inc \dot{Null}(1^0)$
 \TopI$Null(0)$
 \ImpE[\tiny$
 \begin{aligned}
   \langle 0\rangle\inc \dot{Null}(1^0)&\rewrite^* Null(0)\\&\rewrite \top\\
   \langle s(\alpha^0)\rangle\inc \dot{Null}(1^0)&\rewrite^*
   Null(s(\alpha^0))\\&\rewrite \bot
 \end{aligned}$
]$\bot$
\ImpI[(i)]$0=s(\alpha^0)\imp\bot$
\AllI$\forall\alpha^0.~\neg~0=s(\alpha^0)$
\end{prooftree}

Let $E_{\text{\ref{eq:s_inv}}} $ be $\doteq(S^0(\alpha^0), pred(1^0))$,
(\ref{eq:s_inv}) is proved by
\begin{prooftree}
\Axiomm[i]$s(\alpha^0)=s(\beta^0)$
\AllE$\langle s(\alpha^0)\rangle \inc E_{\text{\ref{eq:s_inv}}}\imp\langle s(\beta^0)\rangle \inc E_{\text{\ref{eq:s_inv}}}$
  \Axiomm[i]$\langle\alpha^0\rangle\inc p^c$
  \ImpI[(i)]$\langle\alpha^0\rangle\inc p^c\imp\langle\alpha^0\rangle\inc p^c$
  \AllI$\alpha^0 = \alpha^0$
\ImpE$\alpha^0=\beta^0$
\ImpI[(i)]$s(\alpha^0)=s(\beta^0)\imp\alpha^0=\beta^0$
\AllI[2$\times$]$\forall\alpha^0\beta^0.~s(\alpha^0)=s(\beta^0)\imp\alpha^0=\beta^0$
\end{prooftree}

Let $E_{\text{\ref{eq:s_surj}}}\df (\doteq(1^0,S^0(0))\supset\emptyset)\supset\mathcal P^0(\doteq(S^0(1^0),s(1^0)))$.
(\ref{eq:s_surj}) is proved by
\begin{prooftree}
 \Axiomm[ii]$\langle 0\rangle\inc p^c$
  \ImpI[(ii)]$\langle 0\rangle\inc p^c\imp\langle 0\rangle\inc p^c$
\AllI$0=0$
\Axiomm[i]$0=0\imp\bot$
\ImpE$\bot$
\BotE$\exists\beta^0.~\alpha^0=s(\beta^0)$
\ImpI[(i)]$\neg 0=0\imp\exists\beta^0.~\alpha^0=s(\beta^0)$
\Axiomm[iii]$\langle s(y)\rangle\inc p^c$
  \ImpI[(iii)]$\langle s(y)\rangle\inc p^c\imp\langle s(y)\rangle\inc p^c$
\AllI$s(y)=s(y)$
\ExI$\exists\beta^0.~s(y)=s(\beta^0)$
\ImpI$s(y)\inc E_{\text{\ref{eq:s_surj}}}$
\ImpI$y\inc E_{\text{\ref{eq:s_surj}}} \imp s(y)\inc E_{\text{\ref{eq:s_surj}}}$
\AllI$\forall y.~y\inc E_{\text{\ref{eq:s_surj}}} \imp s(y)\inc E_{\text{\ref{eq:s_surj}}}$
\AndI$\langle
0\rangle\inc E_{\text{\ref{eq:s_surj}}}\wedge\forall y.~\langle y\rangle\inc E_{\text{\ref{eq:s_surj}}}\imp\langle
s(y)\rangle\inc E_{\text{\ref{eq:s_surj}}}$
\OrI$\alpha^0\inc E_{\text{\ref{eq:s_surj}}}$
\AllI$\forall\alpha^0.~(\neg~\alpha^0=0)\imp\exists\beta^0.~\alpha^0=s(\beta^0)
$
\end{prooftree}

(\ref{eq:plus_0}) to (\ref{eq:mult_s}) are easy to prove using the
arithmetical rules and the rule for $=$.

(\ref{eq:ind}) has the following proof:
\begin{prooftree}
  \Axiomm[i]$P(0)$
\Axiomm[ii]$\forall \beta^0.~P(\beta^0)\imp P(s(\beta^0))$
\AndI$\langle 0\rangle\inc E_P^x\wedge\forall
\beta^0.~\langle\beta^0\rangle\inc E_P^x\imp \langle s(\beta^0)\rangle\inc E_P^x$
\OrI[$\langle\alpha^0\rangle\inc E_P^x\rewrite\langle\alpha^0\rangle\inc E_P^x\vee \dots$]$\langle\alpha^0\rangle\inc E_P^x$
\AllI$\forall \alpha^0.~P(\alpha^0)$
\ImpI[$2\times$: (i), (ii)]$P(0)\imp(\forall \beta^0.~P(\beta^0)\imp
P(s(\beta^0)))\imp \forall \alpha^0.~P(\alpha^0)$
\end{prooftree}

(\ref{eq:comp}) has the following proof:\\[.5em]
{\hfill
    \Axiomm[i]$\beta^j\in^j comp^{j+1}(E_A^x)$ \ImpI[(i)]$\beta^j\in^j
    comp^{j+1}(E_A^x)\imp\langle\beta^j\rangle\inc E_A^x$
    \Axiomm[ii]$\langle\beta^j\rangle\inc E_A^x$
    \ImpI[(ii)]$\beta^j\inc E_A^x\imp\langle\beta^j\rangle\in^j
    comp^{j+1}(E_A^x)$ \AndI$\beta^j\in^j
    comp^{j+1}(E_A^x)\Leftrightarrow\langle\beta^j\rangle\inc E_A^x$
    \AllI$\forall\beta^j.~\beta^j\in^j
    comp^{j+1}(E_A^x)\Leftrightarrow A(\beta^j)$ \ExI$\exists
    \alpha^{j+1}.~\forall\beta^j.~\beta^j\in^j
    \alpha^{j+1}\Leftrightarrow A(\beta^j)$
\bottomAlignProof\DisplayProof\hfill\qed}

What we obtain is a conservative extension:
\begin{theorem}
   For all proposition $P$ in the language of $Z_i$, if\quad
   $\Nvdash{}_{\HHA_i} P$ then $Z_i\Nvdash{} P$.
\end{theorem}
\begin{proof} 
  First, we can show, as \citeauthor{arith} do \citeyearpar{arith},
  that adding $pred$, $Null$ and the axioms \eqref{pred_0} to
  \eqref{Null_s} gives a conservative extension. This can be done by
  Skolemizing the proposition $\forall \alpha^0.~\exists
  \beta^0.~(\alpha^0=0\imp \beta^0=0)\wedge(\forall
  \gamma^0.~\alpha^0=s(\gamma^0)\imp \beta^0=\gamma^0)$, which holds
  in first-order arithmetic, and by interpreting $Null(x)$ as $x = 0$.

Then,  we apply the method of \cite{fk:classes06}, which gives a
conservative extension. Finally we skolemize the axioms corresponding to
the comprehension schemata, and thus we obtain a conservative
extension \citep[see][]{dalen89logic}. Then, we have to prove
the equivalence of \eqref{eq:refl} and \eqref{eq:cong_p_c} with
\eqref{eq:eq_def}, which is easy. Finally, we prove that \eqref{eq:ind_c} and \eqref{eq:ind_or} are equivalent.

It can be remarked that the presentation obtained is compatible with
$\HHA_i$, hence the conclusion of the theorem.
\end{proof}

Compared to $\HO_i$, the main issue is that the derivational
complexity of $\HHA_i$ is not polynomially bounded---actually, it does
not even terminate. The non-termination is due to the rule encoding
the induction principle, since it can be proved that the complexity of
$\HHA_i$ without this rule is polynomially bounded. It is not too
surprising, since the real power of arithmetic lies in this
principle. Note that \citet[Remark~1]{arith} propose a terminating
rule to encode the induction principle, but, as stated before, proof
length is not kept. 

\citet{poincare02science} advocates that everything in first-order
arithmetic but the induction principle should be presented as
computation, because the induction principle represents the only real
deductive axiom of the theory. Following this idea, we want to keep
all rewrite rules of $\HHA_i$ excluding the rule for the induction
principle, and present this latter rule in another way. Instead of
using it as an axiom, we can apply the ideas within supernatural
deduction~\citep{Wack} on it. Supernatural deduction consists in
transforming proposition rewrite rules into new inference rules. It
cannot be applied in our case, since $\vee$ cannot be handled by
supernatural deduction. However, it instigates the new inference rule
\begin{prooftree}
  \AxiomC{$\langle 0\rangle \inc \sigma^c$}
  \atoc$\langle \beta^0\rangle \inc \sigma^c$$\langle
  s(\beta^0)\rangle \inc \sigma^c$
\LeftLabel{Ind-i}
\RightLabel{\small
  \begin{tabular}{@{}l}
$\beta^0$ not free in $\langle \tau^0\rangle \inc
  \sigma^c$\\  nor in the assumptions above
\end{tabular}
}
\BinaryInfC{$\langle \tau^0\rangle \inc \sigma^c$}
\end{prooftree}
Proving with this new inference rules is equivalent to proving using
the axiom \eqref{eq:ind_or}. We obtain a first-order proof system for
higher-order arithmetic which is axiom-free, whose proofs can be
checked in polynomial time, and whose proof lengths are the same as in
the axiomatic presentations of higher-order arithmetic.

\begin{note}
  With the rule that we use for arithmetic, we cannot extend the proof
  of strong normalization through reducibility candidates as done by
  \cite{arith}, or through super consistency by
  \cite{dowek06truth}. This still remains an open question whether
proofs of the natural deduction modulo $\HHA_i$ normalizes or not.
\end{note}

\section{Applications to proof-length speed-ups}\label{sec:demo}

Because of Theorem \ref{theo:s3} and Theorem \ref{theo:hhamod}, there is
obviously no proof-length speed-up between $Z_i$, $FZ$ modulo
$\HO_i$ and $\emptyset$ modulo $\HHA_i$. Furthermore, there exists a
speed-up between all these and $Z_{i-1}$, which can be decomposed as follows.

\subsection{Speed-up over compatible theories}\label{sec:demo1}
In this section, we prove that there exists a speed-up between ($FZ$
modulo $\HO_i$) and ($FZ$ and any finite theory compatible with
$\HO_i$). Theorem~\ref{theo:simple} makes it not surprising that, if we
consider $FZ$ plus a finite theory compatible with $\HO_i$, we get a
speed-up with $Z_i$ (or with $FZ$ modulo $\HO_i$). That shows the
interest of using deduction modulo. 
\begin{proposition}\label{prop:speedupWS}
For all $i$, there is an infinite family $\mathcal F$ such that such
that for all finite presentations $\Gamma_i$ compatible with $\HO_i$,
   \begin{enumerate}[\em(1)]
   \item for all $ P\in\mathcal F$, we have $FZ,\Gamma_i\Nvdash{\,} P$
   \item there is a fixed $k\in\mathbb N$ such that for all
     $ P\in\mathcal F$, we have ${FZ\Nvdash{k\text{ steps}}_{\HO_i} P}$
   \item there is no fixed $k\in\mathbb N$ such that for all
     $ P\in\mathcal F$, we have ${FZ,\Gamma_i\Nvdash{k\text{ steps}} P}$
   \end{enumerate}\pagebreak[0]
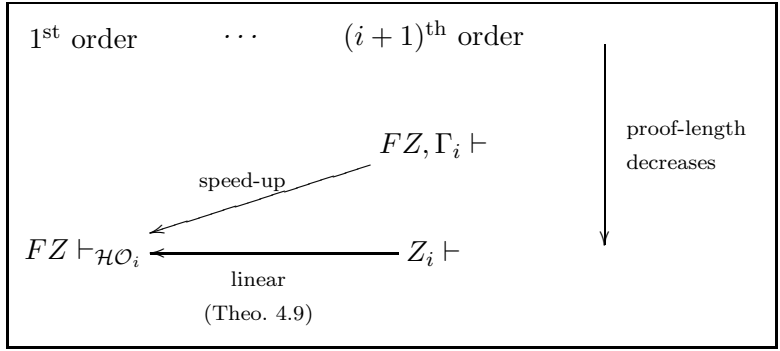
  $$\fbox{\xymatrix{
    \textup{1\ist\ order} & \text{\ldots} & 
    \textup{$(i+1)$\ith\ order} & \ar[dd]^{\textup{\begin{tabular}{l}proof-length\\decreases\end{tabular}}}\\
    &&FZ,\Gamma_i\vdash\ar[dll]_{\textup{speed-up}}\\
    FZ\vdash_{\HO_i} &&
    Z_i\vdash\ar[ll]^{\textup{\begin{tabular}{c}linear\\(Theo.~\ref{theo:s3})\end{tabular}}}&&\\
  }} $$
\end{proposition}
\begin{proof}
  As in the proof of Theorem~\ref{theo:simple}, we first consider the
  standard finite presentation $HO_i$ compatible with $\HO_i$, that
  is, axioms from \eqref{eq:ws_begin} to \eqref{eq:ws_end} and axioms
  \eqref{eq:comp_c_s}.  Consider the set of propositions\
  corresponding to all instantiations of the comprehension schema
  (Comp$^{i-1}$). In $FZ$ modulo $\HO_i$, these propositions can be proved in
  five steps as done in Fig.~\ref{fig:transc}. Obviously, $Z_{i-1}$ is
  not enough to prove all of them, so that (Comp$^{i-1}_{sk}$) has to be
  used in the proofs in $FZ,HO_i$. Nevertheless, the term of sort $c$
  instantiated in it cannot have a bounded size. Then, the
  decomposition of this term using $HO_i$ cannot be done in a bounded
  number of steps. We then use Proposition~\ref{prop:fin_sim} to
  extend this to any finite presentation compatible with $\HO_i$.
\end{proof}

\subsection{Speed-up due to higher orders}\label{sec:demo2}
It is also possible to get a speed-up between $FZ$ plus any presentation
compatible with $\HO_i$ and $Z_{i-1}$.

\begin{proposition}\label{theo:2}For all $i> 0$, 
there is an infinite family $\mathcal F$ such that for all presentations
  $\Gamma_i$ compatible with $\HO_i$, 
   \begin{enumerate}[(1)]
   \item for all $ P\in\mathcal F$, we have $Z_{i-1}\Nvdash{\,} P$
   \item there is a fixed $k\in\mathbb N$ such that for all
     $ P\in\mathcal F$, we have ${FZ,\Gamma_i\Nvdash{k\text{ steps}} P}$
   \item there is no fixed $k\in\mathbb N$ such that for all
     $ P\in\mathcal F$, we have ${Z_{i-1}\Nvdash{k\text{ steps}} P}$
   \end{enumerate}
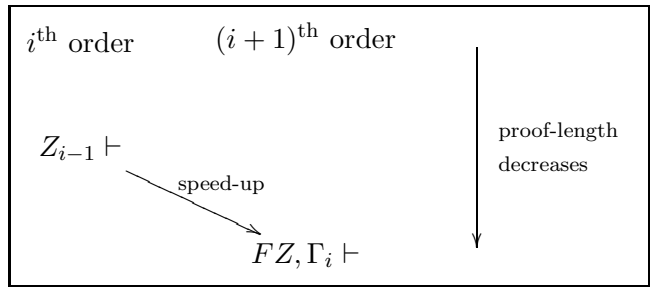
  $$\fbox{\xymatrix{
 \textup{$i$\ith\ order} &
    \textup{$(i+1)$\ith\ order} & \ar[dd]^{\textup{\begin{tabular}{l}proof-length\\decreases\end{tabular}}} \\
    Z_{i-1}\vdash \ar[dr]^{\textup{
          speed-up
    }} \\
   &FZ,\Gamma_i\vdash&
    }} $$
  \end{proposition}
\begin{proof}
If we look at Buss' proof of Theorem \ref{theo:buss}, the infinite
family of propositions\ he use are of the form $P(n)$ where $\forall
n.~P(n)$ can be proved in $Z_i$ whereas in $Z_{i-1}$, $P(n)$ can be
proved, but not with less than $n$ steps. So to get a speed-up it is
sufficient to prove that $\forall n.~P(n)$ can be proved in $FZ$
plus $\Gamma_i$, which is the case because of Theorem
\ref{theo:s3} and \citep[Proposition 1.8]{DHK-TPM-JAR}. We also need Proposition~\ref{prop:NtoH} to show that
if the length of the proofs in $Z_{i-1}\Nvdash{\,}$ was bounded, it would
be the same in $Z_{i-1}\Hvdash{}$, hence a contradiction with
Theorem~\ref{theo:buss}.
\end{proof}

\begin{figure}[b]
  \renewcommand{\arraystretch}{.8}
  $$\xymatrix{
    \text{0\ith\ order} & \text{1\ist\ order} & \text{\ldots} &  \text{$i$\ith\ order} &
    \text{$(i+1)$\ith\ order} & \\
    & Z_0\vdash\ar[ddr]_{\text{speed-up (\hyperlink{hl:theo:buss}{Buss})}} &&&&&
    \ar[dddddd]_{\text{\begin{tabular}{l}proof-length\\decreases\end{tabular}}}\\\\
    &&\ddots\ar[ddr]_{\text{speed-up (\hyperlink{hl:theo:buss}{Buss})}}\\\\
    &&& Z_{i-1}\vdash \ar[dr]^{\text{\begin{tabular}{l}
          speed-up\\(Prop. \ref{theo:2})
        \end{tabular}
    }}\ar[ddr]_(.3){\text{
        speed-up
        (\hyperlink{hl:theo:buss}{Buss})}}  \\
    &&&&FZ,\Gamma_i\vdash\ar[d]^{\text{\begin{tabular}{l}
          speed-up\\(Prop. \ref{prop:speedupWS})
        \end{tabular}
    }}\\
    \vdash_{\HHA_i} & FZ\vdash_{\HO_i} &&&
    Z_i\vdash\ar@/^1.5pc/[llll]^{\text{\begin{tabular}{c}linear\\(Theo.~\ref{theo:hhamod})\end{tabular}}}\ar[lll]_{\text{\begin{tabular}{c}linear\\(Theo.~\ref{theo:s3})\end{tabular}}}&&\\
  } $$
  \caption{Speed-ups in higher-order arithmetic and deduction modulo}\label{fig:sum}
  
\end{figure}
The links between the different systems for higher-order arithmetic
presented in this paper are summarized in Figure~\ref{fig:sum}.

\section{Conclusion and discussion}
In this paper, we have proposed a rigorous framework to study proof
lengths in deduction modulo, by imposing that proofs must be checkable
in polynomial time. We have shown that even with this strict
condition, proofs in deduction modulo can be arbitrarily shorter than
proofs using axiomatizations. We have applied these ideas to study the
length of proofs in higher-order arithmetic. We have encoded higher
orders as a first-order rewrite system, and proved that proofs have
the same length in higher-order arithmetic and in first-order
arithmetic modulo this system. We have also defined a system for
higher-order arithmetic without resorting to any axiom, where proofs can
be checked in polynomial time and have the same length as in the
higher-order axiomatization. All these results open
interesting issues that we discuss below.

The first question that arises from this work is the definition of
what should be considered as a proof. Until recently, automated
theorem provers only answered yes or no (or maybe), and if the prover
was correct, this could be considered as a proof. Of course, the
correction of such provers, often implemented using low-level tricks
to increase the efficiency, is hard to prove. Therefore, many provers
now generate certificates that can be checked in more trustworthy
provers (such as proof assistants like Coq or Isabelle). These
certificates can therefore be considered as proofs, although they may
not contain all the steps that would be included in a usual formal
proof, but only the hints that make it possible to build the formal
proof. This idea is also important for proof-carrying
codes~\citep{necula97:proof}: in this setting, the code of an
application is distributed with a certificate proving its
correctness. The user of the code can therefore check the correctness
using the code and its certificate. It is crucial to have
certificates that are small enough, because they are distributed with
the code, but that can be checked efficiently, because such codes are
often distributed to low-resource systems such as mobile phones. Here
again, a tradeoff has to be found between the details present in the
certificates and the complexity of their checking. Such a tradeoff
could be determined in deduction modulo by choosing what should be
part of the congruence and what should be expressed as axioms. In this
paper, we have advocated that the natural criterion to define what a
proof is, is that it can be feasibly checked. Of course, depending on
the context, this criterion could be relaxed or strengthened.

Another question concerns the role of computation in the speed-ups in
higher-order arithmetic. We have proved, at least to some extend, that
part of these speed-ups originates from the computation (Proposition
\ref{prop:speedupWS}). However, it seems that what really makes proofs
shorter is the fact to be able to reason about higher-order objects,
even if they are encoded by first-order ones (Proposition
\ref{theo:2}). The real point of our results is that it is possible to
use a \emph{finite} first-order encoding while preserving the length
of proofs, at the condition to work modulo some computation. In
general, first-order theorem provers such as Vampire or SPASS only
handle \emph{finitely} presented theories. Note that we have shown in
\citep{burel10embedding} how to integrate deduction modulo into such a
prover.

It could be found inappropriate that rewrite steps are not counted
into the length of the proofs.  Indeed, these steps have to be
performed when searching for the proofs. First, note that it is also
possible to obtain proof-length speed-ups even when counting the
rewrite steps in the length of the proofs, as can be shown by
transposing a result of \citet{bruscoli09complexity} where an
exponential proof-length speed-up is achieved by applying deduction
steps deeply inside propositions
\cite[see][Section~5.2.2]{burel09phd}.  Second, we think that the
speed-ups we obtained should not be considered as cheating, by hiding
part of the proofs in the congruence. This must be thought of as a way
to separate what is deduced and what is computed. To find a proof,
both parts need to be built. To check the proof however, only the
deductive part is necessary, because the rest can be effectively
computed during the verification (hence the need to have a decidable
congruence, even better if it can be decided in polynomial
time). Third, it can also be argued that when the rewrite system is
confluent and polynomially bounded, the rewrite steps are fully
deterministic, so that they do not increase the
proof-search space. Therefore, presenting a theory by means of a rewrite
system instead of a set of axioms can be seen as a way to make proof
search in that theory more deterministic. There are other attempts to
make proof search more deterministic, e.g.\
\citeauthor{andreoli92logic}'s focusing \citeyearpar{andreoli92logic}
in the sequent calculus or \citeauthor{ozan06reducing}'s strategies
for the calculus of structures \citeyearpar{ozan06reducing}, but they
are related to the proof system and not to the theory. Deduction
modulo should be used as a complement to those techniques, when
working in a specific theory. In particular, combining focusing with
deduction modulo leads to what is called superdeduction
\citep{Brauner:2007fk}, as remarked by \citet{houtmann08axiom}.

These results are encouraging indicators that it is as good to work
directly in higher-order logics, as is done in the current interactive
theorem provers, such as Coq ({\tt http://\linebreak[0]coq.inria.fr/}) and
Isabelle/HOL~\citep{Nipkow-Paulson-Wenzel:2002}, or using a
first-order implementation of these logics, as could be done in a
proof assistant based on deduction modulo \citep[or on its sequel
named superdeduction developed by][]{Brauner:2007fk}. It must also be
proved that our results extend to the higher-order systems basing the
interactive provers. This was partly achieved by proving that
functional pure type systems can be encoded in superdeduction in a
manner such that typing inferences in the pure type system are
translated into proofs in superdeduction of the same length
\citep{burel08pts}.  It should also be noticed that in the expression
of HOL in the sequent calculus modulo~\citep{hollambdasigma}, the
length of proofs are preserved too, although it was not highlighted by
the authors.

\section*{Acknowledgement}
The author wishes to thank G.~Dowek, T.~Hardin and C.~Kirchner for
many discussions and comments about earlier versions of this work, as
well as the anonymous referees for their pertinent remarks.

\bibliographystyle{acmtrans}
\bibliography{speedup_LMCS}

\appendix

\section{Translation from
  \texorpdfstring{$Z_i\Nvdash{\,}$}{natural deduction} to
  \texorpdfstring{$Z_i\Hvdash{}$}{schematic systems}}
{\scriptsize\noindent \T{\atoc[\pi]$A$$B$ \ImpI$A\imp B$ \DisplayProof}
  $\df$ \T[A]{\atoc[\pi]$A$$B$\DisplayProof}\bigskip

    \noindent\T{%
      \AxiomC{$\pi_1$}
      \noLine\UnaryInfC{$A$}
  \AxiomC{$\pi_2$}
      \noLine\UnaryInfC{$A\imp B$}
      \ImpE$B$
      \DisplayProof}  $\df$   
    \AxiomC{\T{$\pi_1$}}
    \noLine\UnaryInfC{$A$}
 \AxiomC{\T{$\pi_2$}}
     \noLine\UnaryInfC{$A\imp B$}
   \MP{$B$}
    \DisplayProof\bigskip

\noindent
   \T{%
      \AxiomC{$\pi_1$}
      \noLine\UnaryInfC{$A$}
  \AxiomC{$\pi_2$}
      \noLine\UnaryInfC{$B$}
      \AndI$A\wedge B$
      \DisplayProof}  \\{\flushright$\df$ 
   \AxiomC{\T{$\pi_1$}}
   \noLine\UnaryInfC{$A$}
   \AxiomC{\T{$\pi_2$}}
   \noLine\UnaryInfC{$B$}
   \Axiomm[\ref{eq:K}]$B\imp A\imp B$
   \MP{$A\imp B$}
   \Axiomm[\ref{eq:I}]$A\imp A$
   \Axiomm[\ref{eq:AndR}]$\cdots$
   \MP{$(A\imp B)\imp A\imp(A\wedge B)$}
   \MP{$A\imp(A\wedge B)$}
   \MP{$A\wedge B$}
    \DisplayProof\\[2em]}\bigskip

\noindent
 \T{%
      \AxiomC{$\pi$}
      \noLine\UnaryInfC{$A\wedge B$}
      \AndE$A$
      \DisplayProof}  $\df$   
 \AxiomC{\T{$\pi$}}
 \noLine\UnaryInfC{$A\wedge B$}
 \Axiomm[\ref{eq:AndLL}]$A\wedge B\imp A$
 \MP{$A$}
    \DisplayProof\\ and similarly with (\ref{eq:AndLR}) for the other side.\bigskip

\noindent
\T{%
\AxiomC{$\pi$}
      \noLine\UnaryInfC{$A$}
      \OrI$A\vee B$
\DisplayProof}  $\df$   
 \AxiomC{\T{$\pi$}}
 \noLine\UnaryInfC{$A$}
 \Axiomm[\ref{eq:OrRL}]$A\imp (A\vee B)$
 \MP{$A\vee B$}
\DisplayProof\\ and similarly with (\ref{eq:OrRR}) for the other side.\bigskip

\noindent
   \T{%
\AxiomC{$\pi_1$}
      \noLine\UnaryInfC{$A\vee B$}
      \atoc[\pi_2]$A$$C$
      \atoc[\pi_3]$B$$C$
      \OrE$C$
\DisplayProof}  \\{\flushright$\df$   
\AxiomC{\T{$\pi_1$}}
 \noLine\UnaryInfC{$A\vee B$}
 \AxiomC{\T[B]{$\pi_3$}}
 \noLine\UnaryInfC{$B\imp C$}
 \AxiomC{\T[A]{$\pi_2$}}
 \noLine\UnaryInfC{$A\imp C$}
 \Axiomm[\ref{eq:OrL}]$\cdots$
\MP{$(B\imp C)\imp (A\vee B)\imp C$}
\MP{$(A\vee B)\imp C$}
\MP{$C$}
\DisplayProof\\[2em]}\bigskip

\noindent
\T{%
\AxiomC{$\pi$}
      \noLine\UnaryInfC{$\subst Ayx$}
      \AllI$\forall x.~A$
\DisplayProof}  $\df$   
\AxiomC{\T{$\pi$}}
 \noLine\UnaryInfC{$\subst Ayx$}
\Axiomm[\ref{eq:I}]$\subst Ayx\imp\subst Ayx$
\LeftLabel{(\ref{eq:gen})}
   \UnaryInfC{$\subst Ayx\imp\forall x.~A$}
   \MP{$\forall x.~A$}
\DisplayProof\\ Note that the side conditions are satisfied.
\bigskip

\noindent
\T{%
\AxiomC{$\pi$}
      \noLine\UnaryInfC{$\forall x.~A$}
      \AllE$\subst Atx$
\DisplayProof}  $\df$ 
\AxiomC{\T{$\pi$}}
 \noLine\UnaryInfC{$\forall x.~A$}
\Axiomm[\ref{eq:AllL}]$\forall x.~A\imp\subst Atx$
\MP{$\subst Atx$}
\DisplayProof\bigskip

\noindent
    \T{\AxiomC{$\pi$}
    \noLine\UnaryInfC{$\subst Atx$}
    \ExI$\exists x.~A$
\DisplayProof}  $\df$ 
\AxiomC{\T{$\pi$}}
    \noLine\UnaryInfC{$\subst Atx$}
    \Axiomm[\ref{eq:ExR}]$\subst Atx\imp\exists x.~A$
    \MP{$\exists x.~A$}
\DisplayProof\bigskip

\noindent
   \T{\AxiomC{$\pi_1$}
    \noLine\UnaryInfC{$\exists x.~A$}
    \atoc[\pi_2]$\subst Ayx$$B$
    \ExE$B$
    \DisplayProof} $\df$ 
   \AxiomC{\T{$\pi_1$}}
   \noLine\UnaryInfC{$\exists x.~A$}
   \AxiomC{\T[A]{$\pi_2$}}
   \noLine\UnaryInfC{$\subst Ayx\imp B$}
   \LeftLabel{(\ref{eq:part})}
   \UnaryInfC{$(\exists x.~A)\imp B$}
   \MP{$B$}
\DisplayProof\\Note that the side conditions are satisfied.\bigskip

\noindent
\T{\Tnd$A\vee(A\imp\bot)$\DisplayProof}  $\df$ 
\Axiomm[\ref{eq:tnd}]$A\vee(A\imp\bot)$\DisplayProof\bigskip

\noindent
   \T{\AxiomC{$\pi$}
    \noLine\UnaryInfC{$\bot$}
    \BotE$A$
    \DisplayProof}$\df$ 
    \Axiomm[\ref{eq:I}]$A\imp A$
   \AxiomC{\T{$\pi$}}
   \noLine\UnaryInfC{$\bot$}
   \Axiomm[\ref{eq:K}]$\bot\imp(A\imp A)\imp\bot$
   \MP{$(A\imp A)\imp\bot$}
   \Axiomm[\ref{eq:BotL}]$\cdots$
   \MP{$(A\imp A)\imp A$}
   \MP{$A$}
\DisplayProof\bigskip

\noindent
\T{A}  $\df$  A\bigskip

\noindent
\T[A]{\atoc[\pi]$B$$C$
      \ImpI$B\imp C$
      \DisplayProof}  $\df$ 
\T[A]{\AxiomC{\T[B]{$\pi$}}\noLine\UnaryInfC{$B\imp C$}\DisplayProof}
\bigskip

\noindent    \T[A]{
      \atoc[\pi_1]$A$$B$
      \atoc[\pi_2]$A$$B\imp C$
      \ImpE$C$
      \DisplayProof}\\{\flushright $\df$ 
    \AxiomC{\T[A]{$\pi_2$}}
    \noLine\UnaryInfC{$A\imp B\imp C$}
    \Axiomm[\ref{eq:C}]$\cdots
$
    \MP{$ B\imp A\imp C$}
    \AxiomC{\T[A]{$\pi_1$}}
    \noLine\UnaryInfC{$A\imp B$}
    \Axiomm[\ref{eq:T}]$\cdots
$
    \MP{$(B\imp A\imp C)\imp A\imp A\imp C$}
    \MP{$A\imp A\imp C$}
    \Axiomm[\ref{eq:Co}]$\cdots
$
    \MP{$A\imp C$}
\DisplayProof\\[2em]}\bigskip

\noindent
    \T[A]{
      \atoc[\pi_1]$A$$B$
      \atoc[\pi_2]$A$$C$
      \AndI$B\wedge C$
      \DisplayProof}\\{\flushright $\df$ 
   \AxiomC{\T[A]{$\pi_2$}}
    \noLine\UnaryInfC{$A\imp C$}
  \AxiomC{\T[A]{$\pi_1$}}
    \noLine\UnaryInfC{$A\imp B$}
    \Axiomm[\ref{eq:AndR}]$(A\imp B)\imp(A\imp C)\imp A\imp(B\wedge C)$
\MP{$(A\imp C)\imp A\imp(B\wedge C)$}
\MP{$A\imp(B\wedge C)$}\DisplayProof\\[2em]}
\bigskip

\noindent 
    \T[A]{
      \atoc[\pi]$A$$B\wedge C$
      \AndE$B$
      \DisplayProof} $\df$ 
    \Axiomm[\ref{eq:AndLL}]$(B\wedge C)\imp B$
  \AxiomC{\T[A]{$\pi$}}
    \noLine\UnaryInfC{$A\imp (B\wedge C)$}
    \Axiomm[\ref{eq:T}]$\cdots$
\MP{$((B\wedge C)\imp B)\imp A\imp B$}
\MP{$A\imp B$}
\DisplayProof\\and similarly with (\ref{eq:AndLR}) for the other side.
\bigskip

\noindent
    \T[A]{
      \atoc[\pi]$A$$B$
      \OrI$B\vee C$
      \DisplayProof} $\df$ 
    \Axiomm[\ref{eq:OrRL}]$B\imp(B\vee C)$
  \AxiomC{\T[A]{$\pi$}}
    \noLine\UnaryInfC{$A\imp B$}
    \Axiomm[\ref{eq:T}]$\cdots$
\MP{$(B\imp(B\vee C))\imp A\imp (B\vee C)$}
\MP{$A\imp (B\vee C)$}
\DisplayProof\\and similarly with (\ref{eq:OrRR}) for the other side.
\bigskip

\noindent
    \T[A]{
      \atoc[\pi_1]$A$$B\vee C$
      \atoc[\pi_2]$A,B$$D$
      \atoc[\pi_3]$A,C$$D$
      \OrE$D$
      \DisplayProof} $\df$ \\{
 \AxiomC{\T[C]{\begin{tabular}{c}
      \T[A]{$\pi_3$}\\\\$A\imp D$
    \end{tabular}}}
   \noLine\UnaryInfC{$C\imp A\imp D$}
\AxiomC{\T[B]{\begin{tabular}{c}
      \T[A]{$\pi_2$}\\\\$A\imp D$
    \end{tabular}}}
   \noLine\UnaryInfC{$B\imp A\imp D$}
   \Axiomm[\ref{eq:OrL}]$\cdots$
   \MP{$(C\imp A\imp D)\imp (B\vee C)\imp A\imp D$}
\MP{$(B\vee C)\imp A\imp D$}
 \AxiomC{\T[A]{$\pi_1$}}
    \noLine\UnaryInfC{$A\imp (B\vee C)$}
   \Axiomm[\ref{eq:T}]$\cdots$
\MP{$((B\vee C)\imp A\imp D)\imp A\imp A\imp D$}
\kernHyps{5cm}\MP{$\begin{array}[b]{c}A\imp A\imp D\\\vdots\end{array}$}
\insertBetweenHyps{\kern-5cm}   \Axiomm[\ref{eq:Co}]$\cdots$
 \MP{$A\imp D$}
\DisplayProof\\[2em]}\bigskip

\noindent
    \T[A]{
      \atoc[\pi]$A$$\subst Byx$
      \AllI$\forall x.~B$
      \DisplayProof} $\df$ 
 \AxiomC{\T[A]{$\pi$}}
    \noLine\UnaryInfC{$A\imp \subst Byx$}
\LeftLabel{(\ref{eq:gen})}
   \UnaryInfC{$A\imp\forall x.~B$}
\DisplayProof
\\Note that the side conditions are satisfied.\bigskip

\noindent
    \T[A]{
      \atoc[\pi]$A$$\forall x.~B$
      \AllE$\subst Btx$
      \DisplayProof} $\df$ 
   \Axiomm[\ref{eq:AllL}]$(\forall x.~B)\imp \subst Btx$
 \AxiomC{\T[A]{$\pi$}}
    \noLine\UnaryInfC{$A\imp\forall x.~B$}
  \Axiomm[\ref{eq:T}]$
\cdots$
\MP{$((\forall x.~B)\imp
  \subst Btx)\imp A\imp\subst Btx$}
\MP{$A\imp\subst Btx$}
\DisplayProof\bigskip

\noindent
    \T[A]{
      \atoc[\pi]$A$$\subst Btx$
      \ExI$\exists x.~B$
      \DisplayProof} $\df$ 
    \Axiomm[\ref{eq:ExR}]$\subst Btx\imp \exists x.~B$
    \AxiomC{\T[A]{$\pi$}}
    \noLine\UnaryInfC{$A\imp \subst Btx$}
    \Axiomm[\ref{eq:T}]$\cdots
$
    \MP{$(\subst Btx\imp \exists x.~B)\imp A\imp \exists x.~B$}
    \MP{$A\imp \exists x.~B$}
\DisplayProof\bigskip

\noindent
\T[A]{\atoc[\pi_1]$A$$\exists x.~B$
  \atoc[\pi_2]$A,\subst Byx$$C$
  \ExE$C$
  \DisplayProof} \\{\flushright$\df$ 
\AxiomC{\T[B]{\begin{tabular}{c}
      \T[A]{$\pi_2$}\\$A\imp C$
    \end{tabular}}}
   \noLine\UnaryInfC{$\subst Byx\imp A\imp C$}
   \LeftLabel{(\ref{eq:part})}\UnaryInfC{$\exists x.~B\imp A\imp C$}
 \AxiomC{\T[A]{$\pi_1$}}
 \noLine\UnaryInfC{$A\imp \exists x.~B$}
 \Axiomm[\ref{eq:T}]$\ldots$   
 \MP{$(\exists x.~B\imp A\imp C)\imp
 A\imp A\imp C$}
 \MP{$A\imp A\imp C$}
\Axiomm[\ref{eq:Co}]$\cdots
$
 \MP{$A\imp C$}
\DisplayProof\\Note that the side conditions are satisfied.}\bigskip

\noindent
    \T[A]{
      \atoc[\pi]$A$$\bot$
      \BotE$B$
      \DisplayProof} $\df$ 
    \AxiomC{\T[A]{$\pi$}}
 \noLine\UnaryInfC{$A\imp\bot$}
\Axiomm[\ref{eq:BotL}]$(A\imp\bot)\imp A\imp B$
\MP{$A\imp B$}
 \DisplayProof
\bigskip

\noindent
\T[A]{$A$}  $\df$  \Axiomm[\ref{eq:I}]$A\imp A$\DisplayProof\bigskip

\noindent
    \T[A]{\AxiomC{$\pi$}
   \noLine\UnaryInfC{$B$}
      \DisplayProof}  $\df$ 
     \AxiomC{\T{$\pi$}}
    \noLine\UnaryInfC{$B$}
     \Axiomm[\ref{eq:K}]$B\imp A\imp B$
    \MP{$A\imp B$}
    \DisplayProof~~~\begin{tabular}{p{5cm}}
    if the assumption $A$ is not actually used in $\pi$.
    \end{tabular}
    \bigskip}

\noindent

The definition of T$_A$ for $\imp$-i is not
looping, because they are no longer $\imp$-i in
\T[B]{$\pi$}. Nevertheless, this case impose use to define what T$_A$
means for a proof using the inference rules (\ref{eq:gen}) and
(\ref{eq:part}). (The translation of (\ref{eq:mp}) is already defined
because (\ref{eq:mp}) is equal to $\imp$-e.)

{\scriptsize
\noindent\T[A]{
      \atoc[\pi]$A$$B\imp C(\tau)$
      \LeftLabel{(\ref{eq:gen})}\UnaryInfC{$B\imp\forall \alpha.~C(\alpha)$}
      \DisplayProof} $\df$ 
\AxiomC{\T[A]{$\pi$}}
 \noLine\UnaryInfC{$A\imp B\imp C(\tau)$}
\AxiomC{$\varpi_1$}
 \noLine\UnaryInfC{$(A\imp B\imp C(\tau))\imp(A\wedge B)\imp C(\tau)$}
\MP{$(A\wedge B)\imp C(\tau)$}
\LeftLabel{(\ref{eq:gen})}\UnaryInfC{$(A\wedge B)\imp\forall\alpha.~
  C(\alpha)$}
\AxiomC{$\varpi_2$}
 \noLine\UnaryInfC{$\cdots
$}
\MP{$A\imp B\imp\forall\alpha.~ C(\alpha)$}
\DisplayProof}\\
where $\varpi_1$ is any proof of $(A\imp B\imp C)\imp (A\wedge
B)\imp C$, and $\varpi_2$ of $((A\wedge B)\imp C)\imp
   A\imp B\imp C$, using the axiom schemata (\ref{eq:I}) to
   (\ref{eq:AndR}) and the inference rule (\ref{eq:mp}). (Indeed, they
   are valid propositions\ of the intuitionistic propositional logic.)
\medskip

\noindent{\scriptsize    \T[A]{
      \atoc[\pi]$A$$B(\tau)\imp C$
      \LeftLabel{(\ref{eq:part})}\UnaryInfC{$(\exists \alpha.~B(\alpha))\imp C$}
      \DisplayProof}\\{\flushright $\df$ 
\AxiomC{\T[A]{$\pi$}}
 \noLine\UnaryInfC{$A\imp B(\tau)\imp C$}
\Axiomm[\ref{eq:C}]$(A\imp B(\tau)\imp C)\imp B(\tau)\imp A\imp C$
\MP{$B(\tau)\imp A\imp C$}
\LeftLabel{(\ref{eq:part})}\UnaryInfC{$\exists \alpha.~B(\alpha)\imp A\imp C$}
\Axiomm[\ref{eq:C}]$\cdots
$
\MP{$ A\imp \exists \alpha.~B(\alpha)\imp C$} 
\DisplayProof\\}\bigskip
}

\end{document}
